\documentclass[
superscriptaddress,
twocolumn,
nofootinbib,
amsmath,amssymb,
longbibliography,
floatfix,
aps,
prx,
10pt,
]{revtex4-2}
\usepackage[utf8]{inputenc}
\usepackage[dvipsnames, table]{xcolor}
\usepackage{graphicx}
\usepackage{amsmath}
\usepackage{amssymb}
\usepackage{booktabs}
\usepackage{hyperref}
\usepackage[colorinlistoftodos]{todonotes}
\usepackage{caption}
\usepackage{ragged2e}
\usepackage{subcaption}
\usepackage{thmtools}

\usepackage{needspace}
\usepackage{orcidlink}
\usepackage[resetlabels,labeled]{multibib}

\newcites{H}{Hardware specifications}

\newcommand{\ket}[1]{\left |#1 \right \rangle}

\usepackage{multirow}

\usepackage{colortbl}
\usepackage{float}

\begin{document}
\title{Energetic Analysis of Emerging Quantum Communication Protocols}

\author{Raja Yehia \orcidlink{0000-0002-7843-7398}}
\email{raja.yehia@icfo.net}
\affiliation{ICFO - Institut de Ciencies Fotoniques, The Barcelona Institute of Science and Technology, Castelldefels, Spain}

\author{Yoann Piétri \orcidlink{0009-0005-0734-3529}}
\affiliation{Sorbonne Université, CNRS, LIP6, F-75005 Paris, France}

\author{Carlos Pascual-García \orcidlink{0000-0003-0659-7349}}
\affiliation{ICFO - Institut de Ciencies Fotoniques, The Barcelona Institute of Science and Technology, Castelldefels, Spain}

\author{Pascal Lefebvre \orcidlink{0000-0001-8661-1440}}
\affiliation{Sorbonne Université, CNRS, LIP6, F-75005 Paris, France}
\affiliation{KTH Royal Institute of Technology, Stockholm, Sweden}

\author{Federico Centrone \orcidlink{0000-0001-6116-0516}}
\affiliation{ICFO - Institut de Ciencies Fotoniques, The Barcelona Institute of Science and Technology, Castelldefels, Spain}
\affiliation{Universidad de Buenos Aires, Instituto de Física de Buenos Aires (IFIBA),
Ciudad Universitaria, 1428 Buenos Aires, Argentina.}





\begin{abstract}
With the rapid development and early industrialization of quantum technologies, it is of great interest to analyze their overall energy consumption before planning for their wide-scale deployments.  The evaluation of the total energy requirements of quantum networks is a challenging task: different networks require very disparate techniques to create, distribute, manipulate, detect, and process quantum signals. This paper aims to lay the foundations of a framework to model the energy requirements of different quantum technologies and protocols applied to near-term quantum networks. The energy efficiency of communication protocols is defined and a benchmark on the energy consumption of bipartite and multipartite network protocols is presented. An open-source software to estimate the energy consumption of photonic setups is also provided.
\end{abstract}

\maketitle

\section{Introduction}

With the end goal of constructing a global quantum internet \cite{QIavision}, quantum networks are rapidly developing and are already entering a deployment phase. Several national and international initiatives~\cite{MadQCI,EuroQCI} are already in place to establish an architecture allowing for distant nodes to perform quantum cryptographic tasks, such as secret key exchange. In a world with finite resources where energy demands outgrow energy generation, it is therefore crucial to estimate how much energy these networks will consume before their deployment \cite{AlexiaInitiative}. Such a study can reveal limiting factors for future implementations of networks, or even show the energetic advantages of certain quantum technologies over classical ones. Works that estimate the energetic cost of quantum devices are few, but there are indications \cite{MarcoWork,meier2023} that quantum computing may show an energetic advantage before a computational one.  

The classical internet energetic consumption is about 6-12\% of global energy use and contributing to 2-4\% of global carbon emissions. By 2030, it's projected that the energy demand from internet infrastructure could double, largely driven by the rise of AI and increased digital activity~\cite{clasicalproj}. Including the quantum internet in these projections is essential to benchmark the energy consumption of future communication architectures. Comparing classical and quantum communication protocols in terms of energy is a difficult task, mainly because there are no classical equivalents to most quantum networking protocols with the same level of security. For example, a quantum key distribution protocol achieving information-theoretic security cannot adequately be compared to a classical communication protocol providing only computational security.  It is possible, however, to compare different quantum protocols achieving the same functionality. In this study, we focus on this task, while benchmarking the effective energy consumption of near-term quantum communication infrastructures.

This work presents a framework to quantify the energy efficiency of quantum network protocols. The framework provides an estimate of the energy requirements for basic network functionalities, namely Quantum Key Distribution (QKD) and Conference Key Agreement (CKA), which aim to generate a secret private key among end users of a quantum network. These methods and hardware can be applied to most protocols based on photonic implementations, including quantum computing. In particular, the creation and sharing of entangled states among distant parties, believed to be the main goal of most quantum network architectures~\cite{QIRG}, are the building blocks of many other network protocols~\cite{weakcoinflippin,UnfairCoinTossing,Anonymity,QuantumMoney,fedeVoting,centrone2023cost}.

We define the energy efficiency of a network protocol as the ratio of a performance metric, \textit{e.g.} the secret key rate achieved, over the energy consumed to achieve this task. We showcase this metric on the previously mentioned protocols and analyze the performance of networks of $N$ nodes, since their scaling in resources with the number of nodes is non-trivial. Additionally, we provide concrete values on the energy cost of a network generating a target number of relevant data (\textit{e.g.} secret key bits). We take a hardware-dependent approach to compare different implementations of some common protocols using state of the art hardware. 

Taking into account the energy consumption in our benchmarks, instead of solely the rate or the fidelity, gives a unique perspective. For example, our simulations suggest that there exist regimes of parameters for QKD protocols where using detectors with lower detection efficiency but higher energy efficiency results in huge energy savings at the cost of increased execution time. It can also be shown that, depending on the distance, using different wavelengths can result in lower energy consumption for a comparable rate. Another highlight of our results is that, when taking into account classical digital signal processing costs, the apparent energy advantage of continuous variable QKD over discrete variable vanishes. Finally, we identify optimal protocols to achieve multipartite tasks as a function of the number of parties. These results can guide policy making for future quantum technologies that will take part of future communication infrastructures and energy demands.\\


This work is organized as follows: in Section \ref{sec:energymodel} we introduce the model for estimating the energy consumption of photonic setups; the estimation of the energy costs of bipartite QKD protocols is presented in Section \ref{sec:QKDprotocols} and that of multipartite network protocols, in Section \ref{sec:networkcons}. To improve readability, the table of hardware components that were used as reference is deferred to Appendix \ref{app:tableofcomponent} and the description of the modeling of their energy consumption, to Appendix \ref{app:models}. Appendix \ref{app:QKDprotocols} contains the description of the QKD protocols that were simulated, and additional results on time-bin based setups are presented in Appendix \ref{app:timebin}, Appendix \ref{app:measured} displays experimentally measured energy values, and Appendix \ref{app:distribution} touches on the distribution of the power consumption between the users.

\section{Assessing the energy consumption of quantum networks}
~\label{sec:energymodel}
\subsection{Energy efficiency}
 
A quantum network is a graph where the nodes are quantum devices and the links are quantum channels, here considered to be optical fibers. Running a quantum network assumes that all the devices involved are continuously pumping an energy related to the power consumption of the hardware involved, including the ones related to the classical processing. Understanding the hardware involved in quantum network architecture is then crucial to estimate its energy consumption. However, if a certain hardware allows to double the performance but multiply the energy demand by three, one might prefer sticking with other devices. To get an idea of the efficiency of a network, it is thus important to relate the energy consumed to its performance. In this work, we introduce the \textit{energy efficiency} (EE) of a network protocol, as per similar works in classical networks~\cite{classnetwork}. It is defined by the ratio of a given performance metric over the power consumption of the hardware.

Computing the performance of a quantum network protocol can be done in several ways. In the case of bipartite QKD networks, which are the focus of the first half of this paper, the main metric is given by the achieved secret key rate. The energy efficiency of bipartite QKD networks reads:
\begin{equation}
\label{eq:energyefficiency}
    {\rm EE} [\rm bits/Joule] = \frac{\textrm{Secret key rate [bit/s]}}{ \textrm{Power consumption [Joule/s]}}.
\end{equation}

In the case of multipartite quantum networks, the performance can be assessed using different metrics. For example, as a natural extension of the bipartite case, the conference key rate can be used. Other possibilities include the average of entanglement distribution rates between different parties of the network or, in the case of distributed computing networks, the average latency. As a first step, we focus in Section~\ref{sec:networkcons} on conference key rates and all-to-all entanglement generation.

The energy efficiency is a time independent quantity, which makes it an interesting figure of merit for future quantum networks. It measures the balance between the amount of energy consumed by a network architecture and its abilities, in the same way that flops per watt are used to assess the performance of high-performance computers. This metric allows to compare different network architectures, or different protocols achieving the same goal. However, it neglects initialization costs such as the cooling down of some components, since it assumes that a network is already up and running. Note that, in stark contrast with quantum computing~\cite{MarcoWork}, quantum communication metrics are mostly fixed by the setup and environmental parameters such as losses and noise. This means that inputting more energy does not necessarily increase the communication rate.

\subsection{Energy cost of running a network protocol}

To estimate the energy efficiency of a network protocol, one needs to compute the total power consumption of the network, by taking into account all the hardware involved. In this section, we explicit our model to estimate the total energy cost of running a network protocol. This also allows us to provide explicit values of the energy cost, in Watt, of running a specific network protocol on a given architecture, to get a specific objective. 

The energy cost of running a quantum network protocol can be broken down into three main components: a time dependent component associated with the power consumption of the hardware involved in the protocol, a classical component associated with the classical processing operations involved and a fixed component associated with the initialization of the hardware. In a typical network protocol, the hardware involved can be further sorted into three categories. The first is the source, a broad term encompassing all the required components to generate the state of light onto which the information is encoded. The second comprises all manipulations required in the protocol including, for example, changes in polarization applied through modulators, phase shifters, and so on. The third is the detection, where the optical signal is measured.\newpage

We define the total energy cost to run a protocol $\pi$ as the time dependent function $E_{\pi}(t)$, composed of:

\begin{equation}
\label{eq:generalmodel}
\begin{split}
E_{\pi}(t) &= \sum_{i\in\mathcal{H}_{\pi}} E_i(t) + C_\pi + E_\pi(0)\\
        &=\sum^{n_S}_iS^i_{\pi}(t)+\sum^{n_M}_kM^k_{\pi}(t) +\sum^{n_D}_jD^j_{\pi}(t)\\&\hspace{0.5cm}+C_\pi + E_\pi(0).
\end{split}
\end{equation}

where $\mathcal{H}_{\pi}$ regroups all the hardware elements involved in the protocol, divided further into $S^i_{\pi}(t)$, $M^k_{\pi}(t)$, and $D^j_{\pi}(t)$  which represent the energy contribution at time $t$ from the $i$th, $k$th and $j$th components involved in the source, the manipulation and the detection, respectively. These specifications become relevant when considering large networks, since the number of sources $n_S$, manipulations $n_M$ and detections $n_D$ might all scale differently with the number of users. Each of these families of elements is expanded in Appendix \ref{app:models}.  In Table \ref{tab:tableofeverything}, some values of power consumption are shown for different types of devices that are used in the following sections.  We also include the energy cost of initializing the components $E_{\pi}(0)$, in particular the cost of cooling down the detectors to cryogenic temperatures.

On top of optical components, quantum network protocols typically include some classical computing elements that increase the energy cost, and those are denoted as $C_\pi$. Two major families of contributions are included in this term. The first one consists of the classical controls of quantum components during the time of the protocol. Here, the costs of different components that are commonly involved in communication protocols are considered, such as time taggers to record timing events. All other classical costs required during a protocol are modeled by considering that an active computer is present at each node involved in the protocol. This covers classical communications and memory costs, or classical sub-protocols such as coin flipping. 

The second family of contributions is the energy required for post-processing, which refers to the classical algorithms applied to the outcome of quantum processes. This cost is harder to estimate as it depends on various parameters such as the desired level of security, the classical computing architecture, and the choice of post-processing functions that are usually not standardized. However, this post-processing cost can prove to be non-trivial for some of the protocols studied in this work, especially when considering digital signal processing. This particular cost is discussed in Section~\ref{sec:costDSP}.

Eq. \ref{eq:generalmodel} represents the energy consumed by the hardware of a setup after a running time $t$. The time $t$ required to achieve a certain objective, \textit{e.g.} a certain number of secret key bits in the case of QKD networks, can be computed using device specifications and network parameters such as the distance between the parties. It can then be used  as an input in Eq. \ref{eq:generalmodel} to get the total energy cost of achieving a certain objective.

\section{Energy study of bipartite protocols: Quantum Key Distribution}
\label{sec:QKDprotocols}

In this section, we investigate the energetic expenditure of different widespread QKD protocols. Specifically, we use the previously defined energy metrics on the well-known BB84 protocol~\cite{BB84,BB84coherent1},its entanglement-based variant, the E91 protocol \cite{Ekert,BBM92}, as well as a measurement-device-independent QKD (MDI-QKD) scheme, which was developed to alleviate security assumptions on the detection devices~\cite{Makarov2006Detector,Lydersen2010Detectors,Lucamarini_2018,chen2021twinfield}. Then, we expand the analysis to include Continuous Variable QKD (CV-QKD) protocols~\cite{zhang2024continuousvariable} based on coherent states. Of particular interest are the Gaussian-modulated protocol GG02 \cite{grosshans2002continuous} as well as a discrete-modulated approach known as QPSK \cite{ghorai2019asymptotic,denys2021explicit}. This is concluded with a discussion on the cost of the classical post-processing in QKD protocols. All of these protocols are detailed in Appendix~\ref{app:QKDprotocols}, as well as the model employed to estimate their energy consumption.

\subsection{Secret key rate and energy}

The energy efficiency for bipartite QKD protocol is defined in Eq.~\ref{eq:energyefficiency} as the ratio between the secret key rate $K_{\pi}$ achieved using a specific set of hardware component and the power they consumes. The total power consumption is simply the sum of the electrical powers of the devices involved in the protocol, given in Table~\ref{tab:tableofeverything}. $K_{\pi}$ is given by the specification of the hardware used as well as the distance between the parties. For discrete-variable protocols, $K_{\pi}$ is derived from the \textit{raw key rate} $R_\mathrm{DV}$ in bit per channel use and is given by the following formula:
\begin{equation}
\label{eq:generalRate}
 R_\mathrm{DV} =   \left(\prod_i ^{n_S} \mu_i\right ) \left (\prod_j p_j\right) \left (\prod_k 10^{-\eta_k L_k/10} \right),
\end{equation}
where $\mu_i$ is the probability of photon emission of source $i$. For each hardware element $j$ involved, the probabilities $p_j$ represent their different efficiencies, such as detection efficiencies, coupling efficiencies, and so on. Lastly, $L_k$ is the length of the optical fibers $k$ that light should go through with a loss per kilometer of $\eta_k$. 

Going from raw key to secret key in QKD protocols consists of classical rounds of communication between parties, which use part of the generated bits to assess and amplify the privacy of the rest of the bits. While it is a possibility that the energy cost associated to these classical operations actually dominates the cost of QKD, it is possible to compare several QKD protocols between them by making the assumption that their classical cost is the same\footnote{It is probable that for the same amount of bits in the final key register, the energy cost of privacy amplification for different protocols amounts to comparable values. Error correction on the other hand might lead to very different values in comparing DV and CV protocols. Estimating the energy consumption difference between DV and CV is non-trivial.}. In the main text, we investigate mainly the asymptotic secret key rate, which provides a lower bound on the energy cost. However, the analysis for the finite-size case in CV-QKD is done in appendix~\ref{app:fse}. 

Realistic implementations of QKD protocols always display some noise, quantified by the Qubit Error Rate (QBER) in DV-QKD and by the excess noise in CV-QKD. Estimating those noises is crucial to quantify the number of secret bits that can be extracted from a string of shared bits. Appendix \ref{app:QKDprotocols} explains how to derive the secret key rates $K_{\pi}$ for different protocols and different noise. Note again that the rate is mostly fixed by the setup and environmental parameters such as losses and noise and that inputting more energy does not necessarily increase the rate.

In the case of QKD protocols, the secret key rate is the key performance metric to estimate both the energy efficiency and the concrete energy cost of creating of $N_{\mathrm{target}}$ secret key bits between two parties using protocol $\pi$. It allows us to estimate the the execution time of the protocol $t_{\pi}^{\rm target}$, given by
\begin{equation}
   t_{\pi}^{\rm target}= \frac{N_{\rm target}}{r_{\mathrm{source}}K_{\pi}},
\end{equation}

 where $r_{\mathrm{source}}$ is the repetition rate, and $K_{\pi}$ the secret key rate of the protocol, in bit per channel use. The energy cost $E^{N_{\mathrm{target}}}_{\pi}$ can then be derived from Equation \ref{eq:generalmodel}:

\begin{equation}
\label{eq:QKDenergycost}
    E^{N_{\mathrm{target}}}_{\pi} = E_{\pi}(0) + C_{\pi}^{N_{\rm target}} + t_{\pi}^{\rm target}\sum_{i\in\mathcal{H}_{\pi}} P_i,
\end{equation}

 $\mathcal{H}_{\pi}$ are the overall hardware elements of the protocol (including the source, the manipulation, and the detection) and $P_i$ the power of the hardware $i$ (assuming a constant consumption during the execution of the protocol). Finally, $C_{\pi}^{N_{\rm target}}$ is the cost of classical computing elements, that depends on the number of target bits.

It is interesting to note that Eq.~\ref{eq:generalmodel} can also be used to compute the energy consumed by the continuous use of a protocol for a given period of time. This is particularly relevant in the case of QKD, since these protocols can be used to continuously generate key between users until they are used.\\


\subsection{Results}
To simplify the model for discrete variables protocols, sources are considered as black boxes emitting photonic states at rate $r_{\rm source}$ with the same probability $\mu = 0.01$. This encompasses weak coherent sources with low mean photon number per pulse\footnote{A very low photon number per pulse is assumed, which is always true for cryptographic protocols.}, and SPDC based sources that generate entangled states. All the parameters used in the simulation are summarized in Table \ref{tab:baselineparameters}. The values for the initial energy cost $E_0$ and the power consumption of different hardware components can be found in Table \ref{tab:tableofeverything}. The details of how each component is modeled can be found in Appendix \ref{app:models} while the protocols are presented in Appendix \ref{app:QKDprotocols}. \\

 For readability, additional results are shown only in the appendices, in particular a study with time-bin encoding in Appendix \ref{app:timebin}, a study comparing theoretical values to real world measurements of the energy consumption of different pieces of hardware in Appendix \ref{app:measured}, and a study of the distribution of the power consumption between parties in Appendix \ref{app:distribution}. Interested readers are also invited to use the open-source library \cite{github}, which was developed specifically for this work, with their own components and protocols to estimate the energy cost of the experiments.

\subsubsection{Comparison of DV-QKD protocols}
Three different DV-QKD protocols are compared: BB84, E91 and MDI-QKD. Their description can be found in Appendix \ref{app:QKDprotocols}. 

In Figure \ref{fig:EE}, the energy efficiency of these protocols is compared. To obtain a fair comparison, the same set of Superconducting Nanowire Single-Photon Detector (SNSPDs) is considered for all three protocols. It is to be noted that the lasers used in BB84 and MDI-QKD are the same, but the laser used in the SPDC source for entanglement-based QKD works at a different wavelength. The energy cost $E^{\textrm{1 Petabit}}$ of producing $10^{15}$ secret key bits with the same protocols is shown in the inset.

\begin{figure*}
    \centering
    \includegraphics{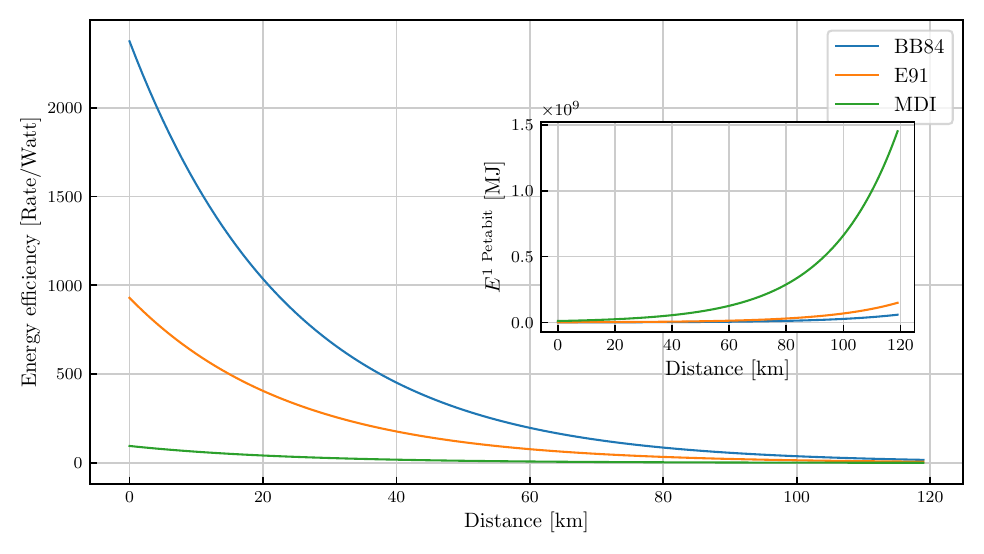}
    \caption{\raggedright\justifying Main plot: Comparison of the energy efficiency of three DV-QKD protocols. Inset: Energy required to produce 1 Petabit of secret key with these protocols.}
    \label{fig:EE}
\end{figure*}

From this plot, it is clear that, for all distance regimes and using comparable hardware, BB84 is the most energy efficient QKD protocol, followed by E91 and then by MDI-QKD. As the ratio of the secret key rate with the power consumed, the energy efficiency directly reflects the high key rate of current BB84 implementations. With the dominating influence of detection apparatus in terms of energy consumption (see Table \ref{tab:tableofeverything} and Appendix \ref{app:distribution}), one could have expected E91 to be the worst performer due to using more detection stations. However, key rates orders of magnitude lower in MDI-QKD make it the most energy consuming, as shown in Table \ref{tab:EEexplained}. Importantly, some qualitative security advantages present in some protocols, like device independence, are not taken into account through this metric. As the success rate of MDI-QKD protocols improves with the development of more efficient quantum memories, adding new components to take into account, this comparison between entanglement based protocols and MDI-QKD protocols might evolve.

\begin{table}[!h]
    \centering
    \begin{ruledtabular}
    \begin{tabular}{ccc}
         Protocol & Power (W) & Secret key rate (kbit/s) \\ \midrule
         BB84 &  2416 & 1092.734  \\
         E91 & 5277 & 934.287 \\
         MDIQKD & 2570 & 46.714 \\ 
    \end{tabular}
    \end{ruledtabular}
    \caption{ \raggedright\justifying  Energy consumption and secret key rate at a distance of 40 km of the three DV-QKD protocols.}
    \label{tab:EEexplained}
\end{table}

It should be noted that the energy efficiency  does not encompass the initialization costs of the devices. It focuses on the energy consumption of a running network, where a specific amount of power is necessary to achieve a given rate. For example, in a BB84 setup using modern cryogenic based SNSPDs shown in table \ref{tab:tableofeverything}, the energy cost of initializing the hardware (in particular waiting 6 h to cool down the detector) is $E_{\rm BB84}(0) = 32$ MJ. To absorb this cost, \textit{i.e.} to have the running setup consume as much as the initialization cost, this takes approximately 3h45. During this time, at a distance of say 40 km, approximately 10 Gbit of secret key can be created. For larger $N_{\mathrm{target}}$, for example 1 Petabit as shown in the inset of Fig.\ref{fig:EE}, the initialization costs are  completely absorbed in the running cost of the setup. 

\begin{figure}
    \centering
    \includegraphics{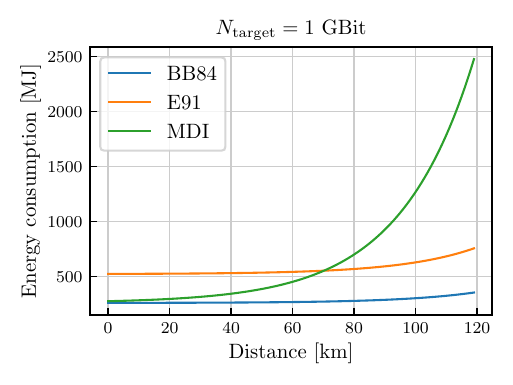}
    \caption{ \raggedright\justifying  Energy required to distill 1 Gbit of secret key using different choices of DV-QKD protocol.}
    \label{fig:DVprotocols}
\end{figure}

Taking into account the initialization costs can lead to a different choice of protocol for a generating a low number of secret bits. In Figure \ref{fig:DVprotocols}, the energy cost $E^{\rm 1 Gbit}$ of the three different DV-QKD protocols producing 1 Gbit $= 10^9$ bits of secret key.  For all distances, BB84 remains the protocol with the smallest energy consumption due to its high rate and the fact that it involves fewer components. However, because they include two detectors, the entanglement-based protocol turns out to be the consuming more energy than the MDI-QKD protocol for distances under 70 km. After this distance, the lower success rate of MDI-QKD protocols makes it more energy consuming than the other two options. 

These results show different insights provided by an energy study of network protocols. Beyond the comparison between QKD protocols, it is also interesting to study how to optimize the hardware for a given protocol. As we show in the next section, different usage of the same protocol might lead to a different choice of hardware to optimize the energy efficiency.

\subsubsection{BB84 Hardware comparison}

\begin{figure}[!ht]
\centering
    \centering
    \includegraphics{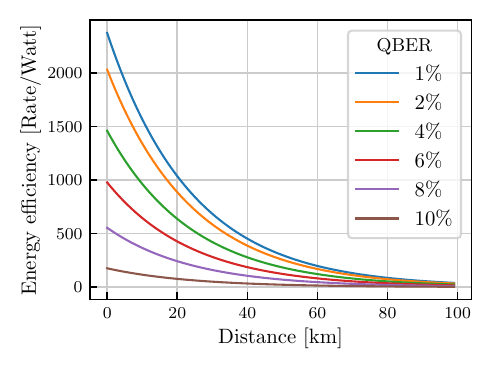}
    \caption{\raggedright\justifying Energy efficiency of a polarization WCP-based BB84 setup as a function of the distance, for different QBER.}
    \label{fig:BB84QBER}
\end{figure}

The most common implementation of BB84 involves a weak coherent state source with the probability of having one photon in a pulse given by $\mu=0.01$, emitting at 1550 nm. The photons are then coupled into a fiber and detected with SNSPDs of efficiency $p_{det}=0.95$. In Figure \ref{fig:BB84QBER}, the energy efficiency of BB84 is shown for different levels of noise, or QBER. The relation between the level of noise in the setup and the energy consumption is not straightforward. It depends on  various passive components or alignment and is not a linear function of the energy, or of the distance between the parties. It can be optimized for a given setup without significantly changing the energy consumed. As an example, a QBER of 2\% was recently reported for different distances in metropolitan implementations of the protocol~\cite{MadQCI}. 

\begin{figure}[!h]

    \begin{subfigure}[t]{0.49\textwidth}        
        \centering
        \includegraphics{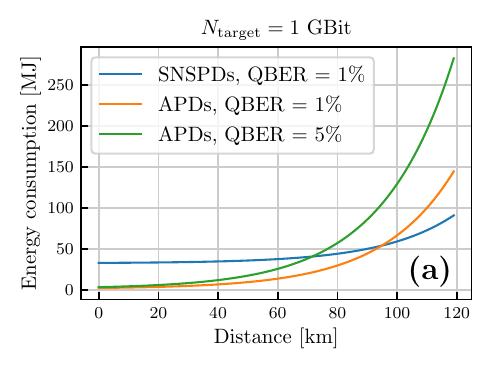}
    \phantomsubcaption
    \label{fig:BB84detect}
    \end{subfigure}
    \hfill
    \begin{subfigure}[t]{0.49\textwidth}
        \centering
        \includegraphics[width = \textwidth]{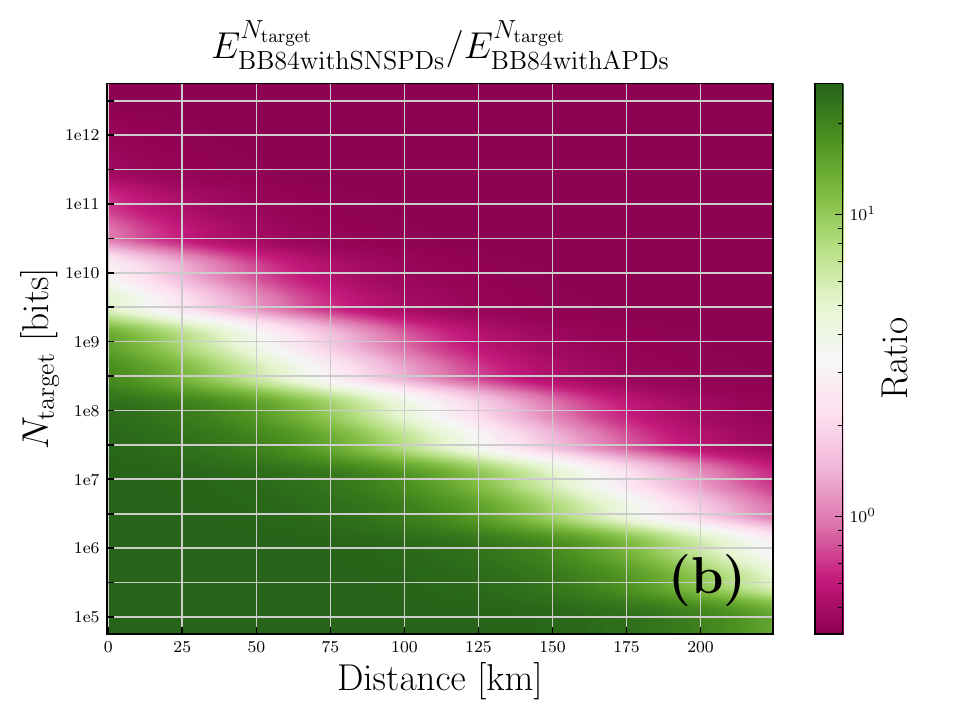}
        \phantomsubcaption
        \label{fig:BB84detect3d}
    \end{subfigure}
    \caption{\raggedright\justifying Influence of the detector choice on the energetic consumption of BB84: \textbf{(a)} Comparison of the energy required to distill 1Gbit of secret key using cryogenic-based SNSPDs and less efficient, potentially noisier, but less energetically costly APDs. \textbf{(b)} Ratio between the energy consumption of BB84 using the two different detectors, for different distances and $N_{\rm target}$. In the green zone, using APDs lead to less energy consumption.}
\end{figure}

BB84 setups require SNSPDs to achieve up to 95\% detection efficiency at $\lambda=1550$ nm. These detectors rely on cryogenic systems that are the main sources of energy consumption in a photonic setup (see Appendix \ref{app:distribution}). In addition to long cooling times, these cryogenic systems require large amounts of energy while they run, as they need to maintain the very low temperatures at which the detectors function. It is then natural to study how this compares to setups using a different detector technology, such as Avalanche Photodiode Detectors (APDs), which require much less starting time and do not require cryogenics, at the cost of a detection efficiency of 25\% at telecom wavelength. In Figure \ref{fig:BB84detect}, we compare the energy consumption of an SNSPD-based setup to one using APDs for the task of generating 1Gbit of secret key, taking into account the initialization costs. We identify a regime of distances for which using APDs consumes less energy than SNSPDs, even considering a higher QBER due to their possible higher dark count rate. This lower energy consumption comes at a cost of execution time : at 20km, using APDs to generate 1Gbit of secret key takes 4 times longer but costs 11 times less energy. This illustrates a trade-off between resource cost and efficiency that can also be observed in other energy studies~\cite{AlexiaInitiative,MarcoWork}. 

The trade-off between the choice of the detector and the energy consumed to generate $N_{\rm target}$ secret key bits depends mostly on the length of this key. As the execution time of the BB84 protocol grows, the cost of cooling down the detector is absorbed and the higher detection efficiency of the SNSPDs compensates for the cost of energy. To illustrate this, in  Fig.~\ref{fig:BB84detect3d} is shown the ratio between the energy consumption of the former over the latter, for different values of the distance and $N_{\rm target}$. The green zone corresponds to a regime where using APDs results in a lower energy consumption. These results can help optimize the choice of the detector apparatus to best adapt to the user's needs. For example if a company wishes to send a new private password once per day to their different offices within a metropolitan radius, they could use BB84 more efficiently using APDs and relaxing the constraints of having to maintain a cryostat.

\begin{figure}[!h]
    \begin{subfigure}[t]{0.49\textwidth}
        \centering
        \includegraphics[width = \textwidth]{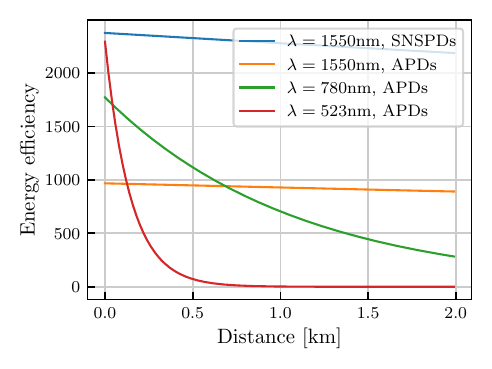}
        \phantomsubcaption
        \label{fig:EEBB84wavelength}
    \end{subfigure}
    \caption{\raggedright\justifying Energy efficiency of BB84 using different choices of wavelengths $\lambda$.}
\end{figure}

To obtain higher detection efficiency without using cryogenic-based detectors, working at other wavelengths can be interesting. Besides the telecom range around $\lambda=1550$ nm, typical wavelength choices are near infrared ($\lambda=780$-800 nm) and visible ($\lambda=523$-532 nm). The critical difference between the wavelengths is the transmissivity of the fiber, as can be seen in Table \ref{tab:baselineparameters}. The energy efficiency of BB84 using different wavelengths is shown in in Fig~\ref{fig:EEBB84wavelength}. It shows that using SNSPDs at 1550nm is still more efficient than using other wavelengths, at the expense of having to maintain cryogenic temperatures. If one does not have the space or resources to use cryostats, it is interesting to note that for hundreds of meter distances, using infrared or visible wavelength can result in a lower energy consumption than the standard telecom wavelength. Since APDs have a higher detection efficiency at lower wavelengths, the time to generate a key is also reduced.

The results of this section hint that in a full-scale QKD network, it might be useful to consider using different wavelengths depending on the distance between all the parties involved and the objective task, to optimize both the overall energy consumption of the network and the key generation rate.

\subsubsection{Continuous Variable QKD} 
\label{SubSub:CVQKD}
The study in this section pertains to CV-QKD protocols, but could be adapted to coherent classical communication protocols, as they use similar hardware. Indeed, the usual setup is based on standard telecom technologies, \textit{i.e.}, at a wavelength of $1550$ nm, and Balanced Homodyne Detectors (BHDs) with typical efficiencies above $p_{det} = 0.7$, and electronic noise around $V_{el}=0.005$ SNU, in agreement with recent advances in the field \cite{pietri2023,Bruynsteen21}. Other parameters used for this section can be found in Table \ref{tab:baselineparameters}. In particular, we choose a higher source rate $r_{\rm source} = 100$ MHz than in the case of DV-protocol, which is representative of the latest CV-QKD experiments~\cite{zhang2024continuousvariable}.

\begin{figure}[!h]
\centering
\includegraphics{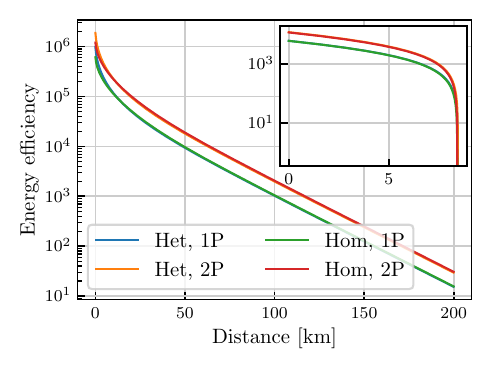}
\caption{\raggedright\justifying  Study of the energy efficiency of CV-QKD protocols. \textit{Main figure :} Energy efficiency of CV-QKD using Gaussian modulation. \textit{Inset:} Energy efficiency for the QPSK scheme. In both cases, the variance of the states was optimized with respect to the distance. In both plots, the orange line is superposed with the red one and the blue line is superposed with the green one.}
\label{fig:CV_Plots}
\end{figure}

The main plot of Figure \ref{fig:CV_Plots} shows the energy efficiency of Gaussian modulated CV-QKD for homodyne- and heterodyne-based detections, as well as single- and double-polarizations. It can be observed that, there is a clear advantage in using a double-polarization scheme since it reduces the overall execution time of the protocol without noticeably increasing the power consumption. Furthermore, the homodyne approach provides a slight improvement in terms of energy for very long distances. The similarity between the homodyne and heterodyne cases is explained by the fact that, while a heterodyne detection measures both quadratures for every round which increases the rate, it then requires two homodyne detection setups which increases the energy cost. Said cost could be reduced by using the RF heterodyne detection scheme, where the two quadratures can be measured with a single balanced detector~\cite{Pietri2024qossthighlymodular}. These two effects were observed to mostly cancel out, with a slight outperformance of the heterodyne setup. This could nonetheless represent an advantage in terms of scalability for networks of many users.

The energy efficiency of the QPSK protocol is shown in the inset of Figure \ref{fig:CV_Plots}. As in the previous case, the double-polarization provides an advantage on the energy efficiency, and both homodyne and heterodyne detections provide an overall similar performance. The experimental setup is identical to the Gaussian modulation scheme with the corresponding detection apparatus but using the QPSK scheme yields a considerably lower key rate, vanishing at high distance. This advantage of the Gaussian distribution might however be affected by the inclusion of the cost of classical post-processing in future implementations, as the digital signal processing and error reconciliation steps may prove to be less energy costly for discrete modulations. Other discrete modulations may also prove more efficient such as Quadrature-Amplitude Modulation, which may make use of a Probabilistic Constellation Shaping~\cite{roumestan2021highratecontinuousvariablequantum}.

\subsection{Classical costs and comparison between CV-QKD and DV-QKD}
\label{sec:costDSP}
Results from Figure \ref{fig:BB84detect} and Figure \ref{fig:CV_Plots} hint at CV-QKD being less energy consuming than any DV-QKD protocols. However, in order to get a meaningful comparison between DV- and CV-QKD protocols, one has also to consider the energetic costs of classical post-processing, which are referred to as \textit{classical costs} in the rest of this section. In addition to being challenging to estimate, they also largely differ from one family of protocols to the other.

We consider the following contributions to the classical costs: signal processing, parameter estimation, secret key rate computations, information reconciliation and privacy amplification. Except for signal processing and information reconciliation, we assume that the same techniques can be used for DV and CV protocols, and hence that the energetic costs are the same. These contributions can thus be ignored in the comparison between the protocols. 

In practice, the biggest difference between CV- and DV-QKD is the digital signal processing (DSP). Indeed, in the DV case, where the photons are detected using single photon detectors, the signal processing is partly done by the time tagger, that has a known energy consumption and is taken into account in the DV setups. On the other side, recent CV-QKD setups (\cite{zhang2024continuousvariable,Pietri2024qossthighlymodular, Jain2022}) are using advanced digital signal processing techniques, which are more costly in energy and cannot be fully realized in real-time at the time of writing of this paper. In this study, we take into account this DSP cost for CV-QKD. Note, however, that the DSP handles a certain number of impairment corrections, also accounting for synchronization (in time, frequency, phase, and sometimes in polarization) which is also needed in DV-QKD experiments. In this sense, we are neglecting some of the DV classical costs, which implies a slight underestimation compared to the cost of CV-QKD. Note also that the information reconciliation cost may differ significantly between CV- and DV-QKD, as error correction for complex variables is more involved than for binary variables. However, estimating the difference between the two costs is not trivial, and is not considered in this analysis.

In CV-QKD, the signal is acquired by the Analog-to-Digital Converter (ADC), and then a series of filters and classical algorithms are applied to recover the symbols. We assume that the cost to recover one symbol from the original signal is a constant and denote it as $\tau_{DSP}$. The energy contribution from signal processing is then given by $\tau_{DSP}$ multiplied by the number of symbols exchanged over the quantum channels: 
\begin{equation}
    E_{DSP} = \tau_{DSP} \frac{N_{\mathrm{target}}}{K_{\mathrm{CV-QKD}}}, 
    \label{eq:dspcost}
\end{equation}
where $N_{\mathrm{target}}$ is the target number of bits in the final secret key and $K_{\mathrm{CV-QKD}}$ is the secret key rate of the CV-QKD protocol (in bit/symbol). This model was validated on the experimental setup of~\cite{Pietri2024qossthighlymodular} but might not be the most general one.

\begin{figure*}[!ht]
    \centering
    \includegraphics{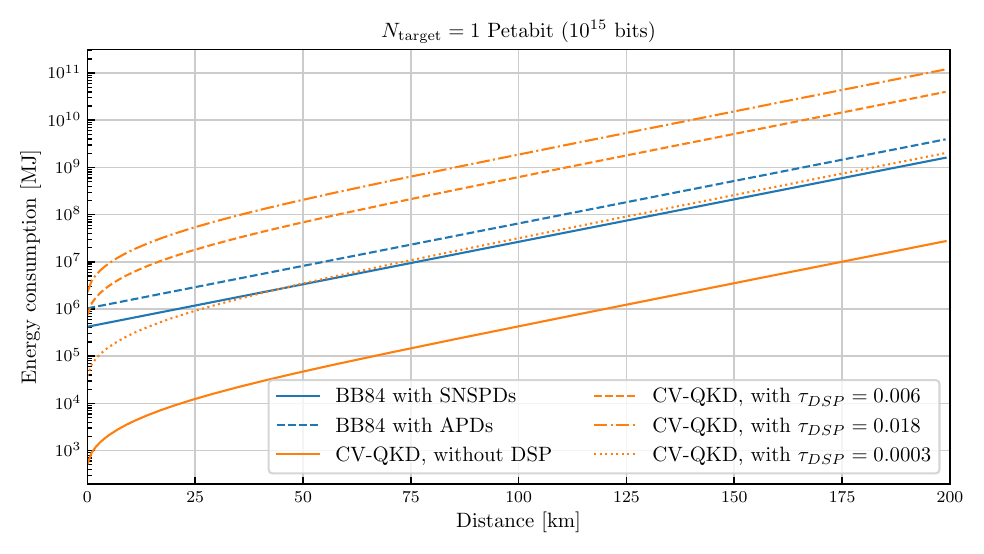}
    \caption{\raggedright\justifying Comparison of the energy consumption of generating 1Petabit of secret key using a DV-QKD BB84 implementation with APDs detectors, SNSPDs detectors and implementations CV-QKD with Gaussian modulation, heterodyne measurement and double polarization, including the classical costs from Digital Signal Processing (DSP).}
    \label{fig:CVVSDV}
\end{figure*}

To get an estimate of $\tau_{DSP}$, the open source QOSST software~\cite{Pietri2024qossthighlymodular} is used as reference, where the DSP runs on a computer during 3 min for 1 million symbols. Assuming a power of 100 W for the computer, this gives an already-achieved value for $\tau_{DSP} = 0.018$ J/symbol. Since the software is written in Python, its running time could be optimized greatly, and a value of 1 min for 1 million symbols could be reached in the near future, which would give a slightly more optimistic value $\tau_{DSP} = 0.006$ J/symbol. Note that since the publication of~\cite{Pietri2024qossthighlymodular}, improvements on the DSP algorithm done partially under the impulsion of the preliminary results of this study allowed the DSP to run in 3 seconds for 1 million symbols on average, giving a $\tau_{DSP} = 0.0003$ J/symbol.

 The energy consumption of generating 1 Petabit of secret key using the BB84 protocol with APDs and with SNSPDs, as well as implementations of Gaussian-modulated CV-QKD without DSP and for the different values of $\tau_{DSP} $ are shown in Figure \ref{fig:CVVSDV}. A large $N_{\mathrm{target}}$ was chosen to limit the effect of the initialization costs. The results reveal that, when neglecting the DSP cost, CV-QKD always outperform BB84. However, when taking into account the DSP cost, DV-QKD can become more efficient than CV-QKD, at it can be seen for the higher DSP costs. Even for the lowest value of DSP, BB84 with SNSPDs start to become more efficient after 50km.

Additionally, note that the DSP cost is independent of the repetition rate $r_{\rm source}$, as seen in Equation \ref{eq:dspcost}.  While increasing the repetition rate of the protocol results in a lower execution time, it does not significantly decrease the overall consumption as the DSP contribution is several orders of magnitude higher than the time contribution. With the same parameters used for the previous simulation, the classical contribution becomes of the same order of magnitude as the time-dependent hardware contribution when $\tau_{DSP}\sim 10^{-6}$ J/symbol. This shows that classical contributions should not be neglected as part of the energetic analysis, and stresses the need for efficient classical post-processing for quantum communication protocols.

Finally at a last remark on Figure~\ref{fig:CVVSDV}, note that this assumes that the noise parameters and reconciliation efficiency for CV-QKD are independent of the distance, which might not be true in practice. As an indication, the current distance records for CV-QKD are 202.81 km~\cite{PhysRevLett.125.010502} using a shared local oscillator and square pulses, and 100 km~\cite{Pi:23, doi:10.1126/sciadv.adi9474} using the newer pulse shaping methods, digital signal processing and local local oscillator. The main limitations come from imperfect phase noise correction, especially in the case of a local local oscillator, and limitations of error reconciliation at low signal-to-noise ratios. For DV-QKD, the distance record for a prepare-and-measure one-way protocol in an optical fiber is 421 km~\cite{PhysRevLett.121.190502} using a three-state protocol with time-bin encoding.

\section{Energy cost of multipartite protocols}
\label{sec:networkcons}

Protocols involving three or more nodes in a quantum network are of particular interest for future deployments of the quantum internet. The energy consumption of quantum devices involved in a network protocol does not scale linearly with the number of users, and therefore it is worth finding optimal configurations for multipartite networks. Firstly, different methods of generation of all-to-all entanglement are compared and secondly the energetic performance of different Conference Key Agreement (CKA) protocols  constructed from DV and CV sources is analyzed.

\subsection{All-to-all entanglement}
All-to-all entanglement generation is the task that consists in creating quantum correlations between the $n$ parties of a network. It can be envisioned as a building block protocol for most multipartite quantum networks: the network continuously generates entanglement between all the parties who can then manipulate and measure their qubits appropriately to reach a desired quantum state for communication or computation.

The most straightforward method to generate multipartite quantum correlations is for one central party to create a state exhibiting genuine multipartite entanglement~\cite{GME} such as the GHZ state~\cite{GHZ}:
\begin{align}
\ket{\mathrm{GHZ}}_n=\frac{1}{\sqrt{2}}(\ket{0}^{\otimes n}+\ket{1}^{\otimes n}).
\end{align}

GHZ states are prime candidates for network applications since they allow the sharing of entanglement between all nodes at once~\cite{Anonymity,fedeVoting,CKAPappa}. A common photonic GHZ state creation setup involves multiple SPDC sources creating Bell pairs that go through a fusion operation (see Appendix \ref{sec:fusion}). Since each source is only emitting $\mu\ll 1$ photon per pulse, the probabilities of simultaneous emission for all the SPDC sources is proportional to $\mu^n$ and the probability of successfully creating a GHZ state of $n$ qubits decreases exponentially with $n$.\\

In Figure \ref{fig:ECGHZ}, an example is shown for $n=4$ for both time and polarization encodings. For polarization encoding, the energy cost is given by:
\begin{equation}
    \label{eq:energyGHZ}
    \begin{split}
    E_{\rm GHZ}(n,t)&= \left \lceil{\frac{n}{2}}\right \rceil  \; S_{\textrm{SPDC}}(t)\\
    &\hspace{0.4cm}+ \left \lfloor{\frac{n-1}{2}}\right \rfloor \; M_{\rm fusion, polarization}(t)\\
    &\hspace{0.6cm}+ n\; D_{\rm SNSPD}(t) + C_{\rm GHZ}(t),
    \end{split}
\end{equation}
where $\left \lceil{x}\right \rceil$ (resp. $\left \lfloor{x}\right \rfloor$) is the integer superior (resp. inferior) or equal to $x$. As before, we model the classical cost $C_{\rm GHZ}(t)$ with a computer in each node involved in the protocol, to perform time-tagging or to record the output.\\

All-to-all entanglement can also be realized through bipartite Bell pairs shared between all pairs of nodes in the network. This architecture requires an SPDC source producing EPR photon pairs between each pair of parties, with each party equipped with single-photon detectors. A round of classical communication after photons are measured allows parties to elevate their bipartite quantum correlations into multipartite ones. The energy associated with such an architecture is given by:
\begin{equation}
    \label{eq:energyAllToAll}
    \begin{split}
    E_{\rm All to all}(n,t)&= \frac{n(n-1)}{2} \; S_{\textrm{SPDC}}(t) + n(n-1) \; M_{\rm polar}(t)\\ &\hspace{0.6cm} + n\; D_{\rm SNSPD}(t) + C_{\rm All to all}(t).
    \end{split}
\end{equation}

\begin{figure}[!ht]
    \centering
    \includegraphics{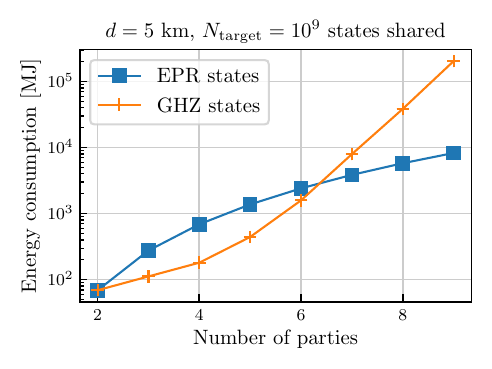}
    \caption{\raggedright\justifying Comparison of the energy required to distribute $10^9$ entangled states between all $n$ nodes of a network as a function of $n$, in log scale. All the parties are separated by an equal distance of $d=5$ km.}
    \label{fig:alltoall}
\end{figure}

While this second method amounts to more hardware involved, the low probability of success of high-order GHZ state creation requires more repetitions and thus longer running times of the hardware. In fact, the best method to share entanglement between all nodes of a network varies with the number of parties, the distance between them and the amount of entangled states to be shared. In Figure~\ref{fig:alltoall}, we show an example of the energy consumption of sharing $10^9$ entangled states between parties separated by 5 km. After $n=6$ parties, the probability of successfully creating a GHZ state becomes so low that it is less energy demanding to use a pair-wise entangled architecture for this task. 

\subsection{Conference key agreement}
Conference key agreement (CKA)~\cite{Murta_2020} is the multipartite extension of QKD, allowing $n$ parties to create and share a common secret key. Similarly to QKD, several protocols achieve CKA, each with different pros and cons in terms of rate, security bound, and energy consumption. \\

This section considers a quantum network of $n$ parties where one central party (Alice) shares states with the others that are denoted as $\{B_i\}_{i=1}^{n-1}$ (Bobs). For simplicity, we consider that all $B_i$ are at equal distances from Alice.

\subsubsection{DV-CKA}

\paragraph{GHZ state implementation}

The most straightforward method to create a secret key between $n$ users of a quantum network is to distribute the qubits of a GHZ state to each party of a network. By measuring their qubits in the appropriate bases, the parties automatically extract a common private bit. A full security proof and description of a GHZ-based CKA protocol can be found in~\cite{Epping_2017}, while an experimental realization can be found in~\cite{ExpCKAGHZ}. The rate of this protocol, which we denote GHZ-CKA, is given by the rate of creating, sharing, and measuring GHZ states that were distributed to the $n$ parties. Assuming that all $n$ parties are at an equal distance $d$ from the source of GHZ states, Equation \ref{eq:generalRate} becomes:

\begin{equation}
    R_{\rm GHZ-CKA}(n) =r_{\rm source} \;  \mu^{\left \lceil{\frac{n}{2}}\right \rceil  }\; p_{\rm coupling}^n \; p_{\rm det}^n \; 10^{n\; \frac{\eta_{\rm fiber} d }{10}},
\end{equation}
 where $d$ is the distance between the source of GHZ states and the parties. As per previous sections, this rate allows the estimation of the time necessary to create a certain objective number of secret key bits between the $n$ parties.  Equation \ref{eq:energyGHZ} can then be used to get the energy cost.\\

\paragraph{Parallel bipartite implementations}

Alternatively, a conference key can be built from $n-1$ bipartite secret keys. Indeed, imagine that Alice shares keys $\{k_i\}_{i=1}^{n-1}$ with each Bob $B_i$. She can choose one of these keys, say $k_1$, to be the conference key and send it securely to each $B_i$. This can be done, for example, using the one-time pad protocol. More explicitly, for each $i\neq 1$, $k_1$ is sent to $B_i$ encoded  with its corresponding key $k_1 \oplus k_i$. $B_i$ can recover $k_1$ by using their key $k_i$.  Two other approaches to CKA are hence considered. In the first one, that we identify as BB84-CKA, Alice performs the BB84 protocol presented in Section \ref{sec:BB84} with each Bob. In the second approach, that we name Bell-CKA, Alice performs the entanglement-based QKD protocol presented in Section \ref{sec:E91} with each other party of the network. The comparison between these protocols is illustrated in Figure \ref{fig:CKAstudyfull}. Other approaches could be envisioned, where a bipartite key is done in parallel between pairs of parties, but those do not compare as directly with the GHZ approach.\\

In the simulations, we assume that all bipartite QKD links work simultaneously, in parallel. Achieving the objective number of shared bits between all nodes therefore requires a time equal to achieving that objective between two nodes only. The total number of sources required in this scenario scales with the number of parties $n$. In the Bell-CKA scenario, the number of detectors scales as $2n$.  The total energy cost is then proportional to the total number of links.

\subsubsection{CV-CKA}

\paragraph{Multipartite Gaussian state implementation}

For continuous variables, a CKA protocol based on the distribution of Gaussian-modulated coherent states is considered, following \cite{ottaviani_modular_2019}. In said protocol, each of the $n$ Bobs individually prepares a copy of an initial state for every round, and all states are sent to a central node where a series of generalized Bell measurements are performed. Namely, a cascade of beam splitters and homodyne detections are applied on the states, such that $n-1$ measurements are performed on the first quadrature and only one measurement is performed on the second. The correlations generated at the beam splitters ensure that all the parties can generate a shared key. In this scenario, the energy model is provided by the following setup: each of the Bobs requires a computer, a laser, an IQ modulator, an MBC, and a DAC. The detection is composed of $n$ homodyne detectors that always measure the same quadrature such that no active phase modulation is necessary, as well as $n-1$ beam splitters which are passive components. The data is then acquired using an ADC where we denote the number of channels as $n_{\rm chan}$. For the subsequent simulations, we consider $n_{\rm chan} = 4$ and the value provided for the ADC in Table \ref{tab:tableofeverything}. Furthermore, we assume that the central detector is linked to one computer that post-processes the measurement outputs. All in all, the energy is given by:
\begin{equation}
    E_\mathrm{CV-CKA}(n,t) = n t P_\mathrm{B,source} + t P_\mathrm{det}(n),
\end{equation}
where:
\begin{align}
    P_\mathrm{B,source} &= P_\mathrm{laser} + P_\mathrm{MBC} + P_\mathrm{DAC} + P_\mathrm{PC}, \\ 
    P_\mathrm{det}(n) &= n P_\mathrm{BHD} + \left\lceil \frac{n}{n_{\rm chan}} \right\rceil  P_\mathrm{ADC} + P_\mathrm{PC} + nP_{\rm laser}.
\end{align}

A brief description of the method to compute the secret key rate of the protocol \cite{ottaviani_modular_2019} is given in Appendix \ref{app:SKRcvcka}. As a concluding remark, each of the Bobs needs to perform post-processing of the measured outcomes, which means that the classical cost increases with the number of users (see Section \ref{sec:costDSP}).\\



\paragraph{Parallel bipartite implementation}

It is possible, as with DV-QKD approaches, to distill a conference key out of $n$ bipartite keys created with CV-QKD protocols. Consider a centralized CV-QKD network, \textit{i.e.}, $n-1$ users (Bobs) connected to a central node (Alice). Each Bob individually distills a key with Alice using the CV-QKD protocol presented in Appendix \ref{app:CVQKD}. More precisely, a Gaussian-modulated CV-QKD protocol is considered, with homodyne measurements and double-polarization. All the parties can then create a common key by mixing the individual, bipartite keys, as explained before. We identify this CKA protocol as the nCV-QKD protocol.

\subsubsection{Simulation results for CKA}

\paragraph{DV-CKA protocols.} For the GHZ-CKA protocol, the central node Alice creates $n$-qubit GHZ-states, keeps one qubit to herself and shares each other one to each $B_i$. For the Bell-CKA protocol, in the same fashion, Alice creates sequentially $n-1$ Bell pairs, keeps one qubit of each pair to herself and sends the other one to each $B_i$. Finally, for the BB84-CKA protocol, Alice sends single photons in the form of weak coherent pulses to each of the $B_i$. Note that the Bell-CKA and the BB84-CKA protocols involve an additional round of classical communication between Alice and each of the Bobs after the quantum key distribution rounds.

\begin{figure}[!h]
    \centering
    \includegraphics{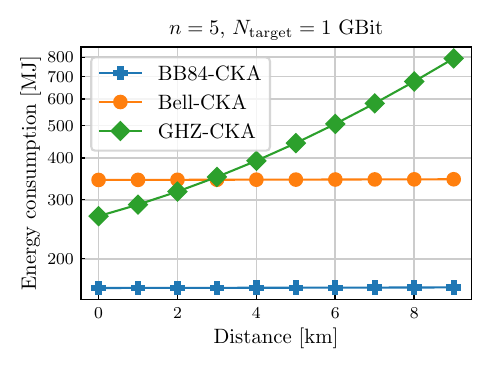}
\caption{\raggedright  \justifying Energy consumption of the three different DV-CKA setups as a function of the distance between the parties and the central node, for a fixed number of parties $n=5$, in log-scale.}
\label{fig:CKAstudyfull}
\end{figure}

\begin{figure*}
    \centering
    \includegraphics{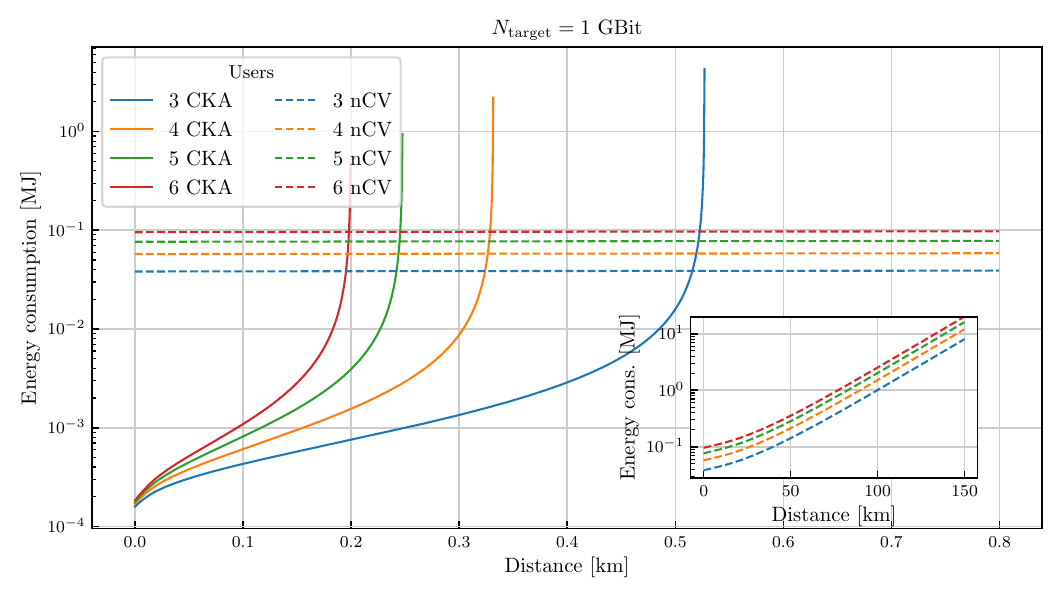}
    \caption{\raggedright\justifying  Energetic analysis of CV-CKA protocols. Main plot: comparison of the energy to distill 1Gbit of key with CV-CKA (plain) and nCV (dashed) protocols as a function of the distance to the central node (short distances only) for different number of users. Inset: Energy consumption required to distill 1Gbit of key with the nCV-QKD protocol (using homodyne measurements and double-polarization) as a function of the distance with the central node, for different numbers of users for larger distances.}
    \label{fig:cvcka}
\end{figure*}

As for all-to-all entanglement generation, the optimal setup depends on the number of parties, the distance with the central node and the target number of secret key bits. In Figure \ref{fig:CKAstudyfull} the energy cost $E^{\textrm{ \rm 1Gbit}}_{\textrm{CKA}}$ required to create 1 Gbit of key is illustrated, using the three aforementioned DV-CKA protocols as a function of the distance with the middle party Alice. The BB84-CKA protocol is always more energy efficient than the other two options. This is due to the high rate and high success probability of the BB84 protocol. Moreover, there is a regime for both the number of parties and for the distance for which the GHZ-CKA protocol is more energy efficient than the Bell-CKA protocol. For longer distances and higher numbers of parties, the probability of a successful generation and detection of a GHZ state decreases to the point where it becomes more efficient to use bipartite communications between Alice and each Bob to accomplish this task. 

As expected, the energy required to create and share GHZ states grows exponentially with the number of parties. Small scale networks can benefit from GHZ-based protocols. Nonetheless, for large number of users, more efficient schemes for GHZ-state creation need to be developed to reasonably envision quantum networks based on multipartite entanglement. 

\paragraph{CV-CKA protocols.} The energy consumption required to create 1 Gbit of secret key using the CV-CKA protocol of~\cite{ottaviani_modular_2019} as a function of distance is shown in Figure \ref{fig:cvcka} for different numbers of parties. It also shows the energy consumption required to create 1 Gbit of secret key using the nCV-QKD protocol for diverse numbers of users $n$, all of them separated by the same distance. A similar energy consumption with respect to the CV-CKA approach is observed with a noticeable improvement in terms of achievable distances (see inset), with the divergences caused by the secret key rate decreasing to zero. However, at the scale of few hundred meters, the CV-CKA approach performs better than nCV-QKD, hinting that such protocols could prove useful in access networks or buildings. Both techniques employ vastly different classical post-processing (in particular, the DSP), such that adding those contributions could modify the results of this simulation in future estimations.

From this section, it could be concluded that the energy consumption of the CV-CKA protocol is three orders of magnitude lower than its DV-counterpart, but with major limitations on the achievable distance since this energetic advantage is valid only over a few hundred meters \cite{ottaviani_modular_2019}. This could therefore be a solution to be considered for a few users within building-scale distances, compared to the DV-CKA protocols presented in previous sections. A complete study should however include the classical post-processing costs of CV-CKA, which are not standardized as of the date of writing this study.

\section{Conclusion}
In this article, we laid the foundations of a framework to analyze the energy properties of quantum network protocols. The energy efficiency, defined as the ratio of some protocol performance metric over the energy consumed, was introduced and showcased. It allows to compare protocols performed on a running quantum network. We were also able to provide concrete values of the energy consumption of current implementations of bipartite or multipartite protocols and optimize on different hardware implementations.

This first insight into the energetic cost of photonic-based quantum communication protocols, including multipartite scenarios, shows that the critical components in discrete variable protocols are cryogenic-based hardware with long cooling times. An interesting highlight of this study is that, for a distance of 20 km, a typical cryogenic-based BB84 setup could generate 1 Gbit of secret key 11 times more efficiently using a room-temperature detector, at the cost of 4 times the temporal requirements. On the other hand, continuous variable protocols are deeply affected by classical post-processing costs. Understanding these differences are critical at this early stage of quantum network development and can direct future development of quantum networks in different ways. While replacing the whole existing infrastructure is not foreseeable, our analysis shows that the installation of new small-scale networks could benefit from thoughtful detector and wavelength choices. 

Due to their probabilistic nature, multipartite protocols based on GHZ states scale worse over long distances or high number of parties than those based on repeating pair-wise DV-QKD, but show competitive regimes at smaller distance and number of parties. Results might evolve by using new hardware such as dedicated integrated photonic platforms to reduce the energy cost. 

While we believe the study conducted in this paper is important and can already guide some policy making decisions, we acknowledge that it does not give a complete view of the cost of a future quantum Internet. One important cost that we do not cover here, which might prove to be the biggest one, is the manufacturing cost. Other effects, such as the usage and the inclusion of satellite links also drastically influence the energy consumption of the whole network. This current work only provides one of the many pieces necessary to make a full life-cycle assessment of a quantum network protocol implementation. 
Future work will extend this study in several directions. This framework could readily be used to estimate the energy consumption of photonic computation architectures, such as~\cite{Quandela}. More consideration must be given to the cost associated with classical post-processing, and in particular to the trade-off between the energy spent in post-processing and the level of security. Since said task is more involved in CV protocols, this might give more insight into which method is more efficient. Furthermore, it is worth considering the authentication cost of the classical channels used by the parties. This is done using pre-shared keys \cite{Dodis2009} or post-quantum cryptography \cite{Dowling2020}, and is a requirement to avoid man-in-the-middle attacks. The cost of these tasks is difficult to estimate because there is no standard method to perform them yet. Finally, the list of hardware available in the accompanying software can be expanded, including for example new components such as quantum memories or solid-state sources. This will contribute to rigorous benchmarking and optimization of the energy consumption of large scale quantum networks with heterogeneous hardware for each nodes and hybrid fiber/free-space links.

\section*{Acknowledgments}
The authors thank Paul Hilaire, Simon Neves and Verena Yacoub for fruitful discussions and inputs, as well as Eleni Diamanti and Alexia Auffèves for their feedback and guidance.\\

The authors acknowledge financial support from the European Union (ERC, ASC-Q, 101040624) and (EQC, 101149233), European Union’s Horizon Europe research and innovation program under the grant agreement No 101114043 (QSNP) and QUCATS, together with Vinnova and Wallenberg Centre for Quantum Technology through the NQCIS project (1011113375), the ERC (AdG CERQUTE, 834266), the PEPR integrated project QCommTestbed ANR-22-PETQ-0011, as well as support from the Government of Spain (Severo Ochoa CEX2019-000910-S and FUNQIP), Fundació Cellex, Fundació Mir-Puig, Generalitat de Catalunya (CERCA program).

Views and opinions expressed are however those of the author(s) only and do not necessarily reflect those of the European Union or the European Research Council. Neither the European Union nor the granting authority can be held responsible for them.







\appendix

\section{Simulation parameters}
\label{app:tableofcomponent}
In Table \ref{tab:baselineparameters} and \ref{tab:tableofeverything}, we present the parameters that we use for all the simulations in this article.

\begin{table*}
    \centering
    \begin{ruledtabular}
\begin{tabular}{lcl}
    Symbol & Value & Description  \\ \midrule
    $r_{\rm source}$ & 80 MHz & Repetition rate of lasers for DV protocols \\
    $\mu$ & 0.01 & Probability of emitting a state from a source\\
    $p_{\rm coupling}$ & 0.9 & Coupling probability into a fiber \\
    $p_{\rm BSM}$ & 0.5 & Success probability of a Bell state measurement \\ \midrule
    $r_{\rm source}$ & 100 MHz & Repetition rate for CV protocols\\
    $V_{el}$ & 0.005 SNU & Electronic noise \\
    $\beta$ & 95\% & Reconciliation efficiency for CV-QKD \\
    $\xi$ & $0.01$ SNU & Excess noise \\ \midrule
     \cellcolor{white}& 0.95 & Detection efficiency of SNSPDs at 1550 nm \\ 
\cellcolor{white}& 0.25 & Detection efficiency of InGaAs-APDs at 1550nm\\
\cellcolor{white}&0.75 & Detection efficiency of Si-APDs at 780nm \\
\cellcolor{white}& 0.5 & Detection efficiency of Si-APDs at 523nm\\
\cellcolor{white}\multirow{-5}{*}{$p_{\rm det}$}& 0.7 & Detection efficiency of the Balanced Homodyne Detector at 1550nm\\ \midrule
\cellcolor{white}& 30 dB/km & Fiber loss coefficient at 532 nm\\
\cellcolor{white}& 4 dB/km & Fiber loss coefficient at 780 nm\\
\cellcolor{white}\multirow{-3}{*}{$\eta_{\rm fiber}$}& 0.18 dB/km & Fiber loss coefficient at 1550 nm\\ 
\end{tabular}
\end{ruledtabular}
    \caption{\raggedright\justifying Baseline simulation parameters. The first section of rows corresponds to DV parameters, while the second refers to CV parameters. The last two are detector efficiencies and fiber loss coefficients for different wavelengths.}
    \label{tab:baselineparameters}
\end{table*}

\begin{table*}
\small
\begin{ruledtabular}
\begin{tabular}{llllllll}
\rowcolor{white} Laser  & $\lambda$ (nm) & $E_0$ (kJ) & Meas. (kJ) & $P$ (W)  & Meas. (W) & Ref.\\ \midrule
Verdi C-Series & 532 & 648 & & 360 & & \citeH[H][]{laser1}\\   \arrayrulecolor{black!10}\hline
Verdi V-Series & 532 & 1620 & 864 & 900 & 480 & \citeH[H][]{verdivseries}\\ \hline
DLC TA pro & 795 &  0 & & 70 & & \citeH[H][]{laser2}\\ \hline
D2547P & 1532 & 0 & & 3 & & \citeH[H][]{laser3}\\ \hline
NKT Koheras Basik X15 & 1550 & 0.12 & 0.126 & 4 & 4.2 & \citeH[H][]{koherasbasikx15} \\ \hline
Mira HP F & 780 &  3240 & & 1800 & & \citeH[H][]{laser4}\\ \hline
SCW 1532-500R  & 1550 & 0  & 2.4 & & & \citeH[H][]{laser5}\\
\arrayrulecolor{black}\midrule

\rowcolor{white} Detector & $\lambda$ (nm)  & $E_0$ (kJ) & Meas. (kJ) & $P$ (W) & Meas. (W) & Ref.\\ \midrule
Si-APD & 523 & 0 & & 45 & & \citeH[H][]{detector5} \\ \arrayrulecolor{black!10}\hline
 Si-APD & 780 & 0 & & 15 & & \citeH[H][]{detector1}  \\ \hline
  InGaAs-APD & 900-1700  &48.3 & 5.04* & 161 & 14 & \citeH[H][]{id220} \\ \hline
 InGaAs-APD & 1532 &1159 & 125.7* & 644 & 64 & \citeH[H][]{detector2} \\ \hline
 SNSPD & 780 & 259200 &  & 3000 & & \citeH[H][]{detector3}  \\ \hline
 SNSPD & 1532 &259200 & 117639* & 3000 & 2735 & \citeH[H][]{detector3}  \\ \hline
SNSPD & 1550 & 32400 & & 1500 & & \citeH[H][]{detector6} \\ \hline
 Balanced detector & 1550 & 0 & & 3 & 6.8 & \citeH[H][]{thorlabspdb480ac}\\
\arrayrulecolor{black} \midrule

\rowcolor{white}Component &    & $E_0$ (kJ) & Meas. (kJ) & $P$ (W) & Meas. (W) & Ref.\\ \midrule
 Computer &    & 9 & 6 & 150 & 100 & \citeH[H][]{component1}\\ \arrayrulecolor{black!10}\hline
 Time tagger & & 0 & & 50 & 22 & \citeH[H][]{component3, component9} \\ \hline
 Motorized Waveplates & & 0.93 & 0.249 & 31 & 8.3 & \citeH[H][]{component2}\\ \hline
 Interferometry  & & 0 & & 200 & & \citeH[H][]{laser3,detector6,component8}\\ \hline
 Modulator (AM) & & 15 & 0.78 & 500 & 26 & \citeH[H][]{component4, MBC-DG-LAB, FI-5682GA}\\ \hline
 Oven (with Controller) & & 9 & 0.54 & 15 & 0.9 & \citeH[H][]{component6}\\ \hline
 Modulator (IQ) & & 0.18 & 0.162 & 6 & 5.4 & \citeH[H][]{mbciqlab}\\ \hline
Polarization Controller & & 0 & & 1.8 & 0.35 & \citeH[H][]{ThorlabsMPC320}\\ \hline
Powermeter & & 0 & & 1 & 0.8 & \citeH[H][]{ThorlabsPM101A}\\ \hline
Optical switch  & & 0 & & 1.8 & 0.35 & \citeH[H][]{thorlabs-osw12}\\ \hline
ADC & & 0 & & 30 & 20 & \citeH[H][]{teledyneadq32}\\ \hline
DAC & & 0 & & 40 & 40 & \citeH[H][]{teledynesdr14tx}\\\arrayrulecolor{black}
\end{tabular}
\end{ruledtabular}

\caption{\raggedright\justifying Example of components frequently used in laboratories, including information about their power usage, startup time, and other interesting data such as central operating wavelengths, detector efficiencies, and reference documentation. Measured values are also indicated (see below for more details on the experimental procedure). }
\label{tab:tableofeverything}
\end{table*}

Theoretical values for $E_0$ are obtained by multiplying the initialization time by their power. Measured $E_0$ is obtained either by multiplying the initialization time by the measured power, or fully measured by following the power consumption in real-time and integrating (marked by an asterisk *).\\

For the measured values in Table \ref{tab:tableofeverything} (columns 5 and 7), the measurements were done in one of the three following ways, depending on its power connection type:
\begin{itemize}
    \item If the device could be plugged through a standard power adaptor to a standard electrical outlet, the measurement was done using a power meter socket~\citeH{powermetersocket} with a maximal load of 3680 W.

    \item If the device was plugged into a standard lab power supply, then the measurement was done by applying the required voltage, and recording the consumed current, and multiplying the two values to get the power.

    \item Finally, if the device was powered through USB, then the consumption was measured by adding a USB adaptor before the power meter socket. This method adds the consumption of the adaptor, but these usually have low power consumption values (USB 3.0 has a maximal output voltage of 5V with a standard requirement of 0.9 A resulting in a typical maximal power of 4.5 W).
\end{itemize}
When possible the power consumption was measured while the device was being used. The measurements for the lasers were done while emitting their optical beam; for the single-photon detectors, while detecting a flow of single photons; for the computers, while they were running the control software of the CV-QKD experiment; for the waveplates, while they were rotating; for the AM modulator, while applying a pulsed signal (for the signal generator) and while locking (for the modulator bias controller); for the oven, while stabilizing a temperature of 25$^\circ$C; for the IQ modulator, while locking the modulator; for the powermeter, while receiving optical power; for the optical switch, while applying switching commands; for the ADC and DAC, while running the CV-QKD experiment (and hence emitting and receiving signals).\\

Additionally, for some equipment with a long initialization time (such as lasers or single photon detectors), the power consumption was recorded during the initialization.

\section{Models for encoding, source, detection and manipulation}
\label{app:models}
This appendix contains details of the models for different source, manipulation and detection elements used in Equation \ref{eq:generalmodel}.
 
\subsection{Choice of encoding}

There are multiple approaches to encode information in a state of light, each with pros and cons, and each involving different hardware to create, control and measure quantum information. For discrete variable (DV) protocols, a simple option is to encode qubits on the polarization of a photon or of a pulse of light. This encoding is of major interest due to easily available passive components, but polarization is susceptible to birefringence which is present in most fiber networks. An alternative is time encoding where two distinct arrival times are defined as the logical zero and one. This type of encoding is also a major contender due to its reliability over long distances, but requires precise control of the phases of the signals due to the necessary use of interferometers. \\

Continuous variable (CV) protocols rely on quadrature encoding and are readily implementable with commercial telecom components. In quantum optical systems, quadratures are the real and imaginary parts of the electromagnetic field~\cite{gast2005802}, which are usually named in-phase and quadrature components in classical telecommunications~\cite{braunstein2005quantum}.  The encoding can be implemented using an IQ modulator and decoded with coherent detection (homodyne or heterodyne). The advantage of this class of protocols is that they can be performed with the same hardware as in the currently deployed telecommunication infrastructure. While they are less resilient to losses compared to their DV counterpart, the higher repetition rate allowed by the detection system usually makes the secure key rate of CV-QKD protocols higher than DV-QKD protocols at metropolitan scale distances~ \cite{zhang2023continuous}. \\

On a different note, this analysis refers to the most popular schemes of encoding, but other choices not considered in this study (such as frequency, spatial and angular orbital momentum encodings) are possible.\\

The choice of encoding in communication protocols influences not only the hardware used but also the performances achieved. In this work, we take a minimalist approach, simplifying setups and protocols to the strict minimum functionalities required to make these protocols work, in order to show general behaviors and to showcase the model in different situations.

\subsection{Sources}

\subsubsection{Weak coherent pulse sources}
Weak coherent pulse sources are a practical and efficient method to obtain single photons probabilistically. The idea consists of attenuating a laser pulse until the probability that a light packet contains more than one photon is low enough that it can be neglected or effectively bounded in security proofs. From a hardware point of view, this requires a laser, passive attenuation and, usually, an amplitude modulator.

\begin{figure}[!ht]
    \centering
        \includegraphics[width=0.45\textwidth]{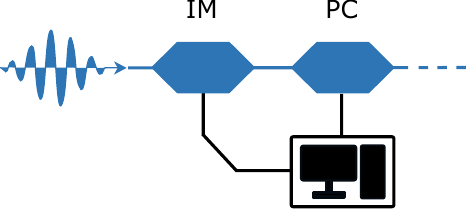}
\caption{\raggedright\justifying Typical source schematic for weak coherent pulses.  The pulses are carved using an intensity modulator, while the polarization is controlled through electro-optical modulation. 
 The lasing power can be attenuated at the laser level, or through passive attenuation.}\label{fig:WCPsource}
\end{figure}

Polarization encoding can be done through electro-optical modulation, which allows automation in the context of large networks \cite{cortes2022bayesian}. This is translated to the following function representing the energy cost:
\begin{align}
S_{\textrm{weak,polarization}}(t) = t(P_{\rm laser} +2 P_{\rm modulator}).
\end{align}

For time encoding, interferometry is done through the use of unbalanced Mach-Zehnder interferometers. These devices require an independent weak laser and a classical detector for stabilization purposes, heating elements for temperature control, and a piezo-type element for phase control between the two arms.  These are included in a general function called $P_{\rm interferometry}$.

\begin{equation}
S_{\rm weak,time}(t) = t(P_{\rm laser} + P_{\rm modulator} + P_{\rm interferometry}).
\end{equation}

\subsubsection{Spontaneous parametric down conversion sources}
Spontaneous Parametric Down Conversion (SPDC) is a popular, cheap and accessible technology that generates light at the single photon level and that, above all, creates correlated photon pairs that are easy to entangle.  While one of the photons of the pair can be ignored when only a single photon is required, many protocols use this second photon as a heralding mechanism \cite{huang2011heralding}. Sources based on SPDC are structurally simple: a pump laser and a non-linear crystal are the minimum requirements to generate single photons.  Note that a resistance heater oven maintains the crystal's temperature. 

\begin{figure}
\centering
    \begin{subfigure}{0.45\textwidth}
        \centering
        \includegraphics[width=0.9\textwidth]{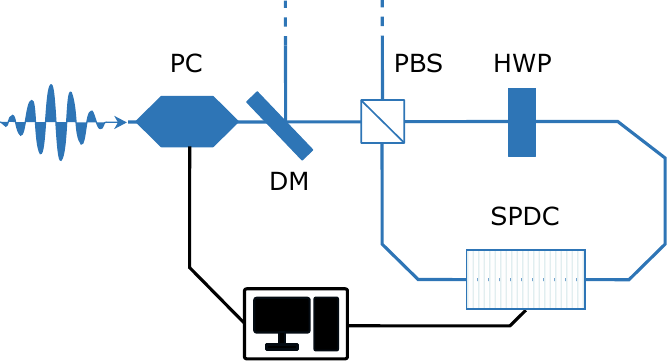}
        \caption{\raggedright\justifying   Source for photon pairs entangled in polarization. }
        \label{fig:SPDCpolsource}
    \end{subfigure}
     \begin{subfigure}{0.45\textwidth}
        \centering
        \includegraphics[width=0.9\textwidth]{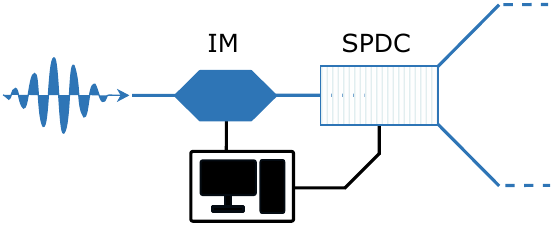}
        \caption{\raggedright\justifying  Source for time entangled photon pairs.}
        \label{fig:SPDCtimesource}
    \end{subfigure}
    \caption{\raggedright\justifying Components involved in SPDC sources: \textbf{a)} a laser light is oriented in the $\ket{H}+\ket{V}$ polarization state through quarter and half-waveplates (Q/HWP) before entering a Sagnac loop consisting of a Polarized Beam Splitter (PBS), a Half-Waveplate (HWP) and a non-linear crystal in which the pump light undergoes Spontaneous Parametric Down Conversion (SPDC).  The created photon pairs exit the loop through the PBS and a Dichroic Mirror (DM). \textbf{b)} Laser light is pulsed into two time bins by an Intensity Modulator (IM) before undergoing SPDC. In both setups, a resistance heater oven maintains the temperature of the crystal.}
\end{figure}

Polarization encoding is done using a Sagnac loop, as shown in Figure \ref{fig:SPDCpolsource}. The energy cost of a polarization SPDC source is thus given by:
\begin{align}
S_{\rm SPDC,polarization}(t)=t(P_{\rm laser} + P_{\rm oven}+P_{\rm modulator}).
\end{align}

For time-bin encoding, interferometry of some type is required to transform a single pulse into two phase-controlled pulses.  This can be done directly with an intensity modulator, although methods exist using Mach-Zehnder interferometers and/or phase modulators.  The simplest setup, shown in Figure \ref{fig:SPDCtimesource}, gives the following energy cost:
\begin{align}
S_{\rm SPDC,time}(t)&= t(P_{\rm laser} + P_{\rm oven} + P_{\rm modulator}).
\end{align}

\subsubsection{Modulated coherent states sources}
\begin{figure}[!ht]
    \centering
        \includegraphics[width=0.45\textwidth]{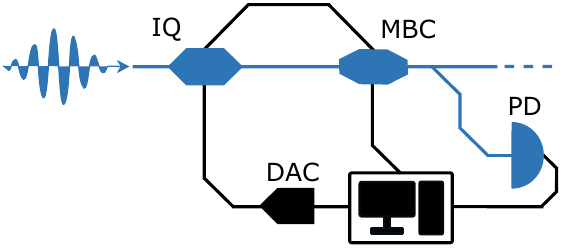}
\caption{\raggedright\justifying Setup for a coherent source of CV states. Information is encoded by sending an electrical signal from a Digital-to-Analog Converter (DAC) to the IQ modulator, effectively displacing coherent states, while the Modulator Bias Controller (MBC) acts as a feedback loop to lock the modulator on its functioning point. A photodiode (PD) is used to monitor the optical power and measure the modulation strength. A realistic setup would also include optical attenuators and other passive optical elements.}\label{fig:CVsource}
\end{figure}

In CV protocols based on modulated coherent states, one needs to generate coherent states and choose their average quadratures, which can be done by using a laser and an IQ modulator~\cite{FundamentalsOfKikuch2016}, which is itself usually composed of 2 Mach-Zehnder interferometers nested in a third one. Such a source would also include passive attenuators to reach the required low modulation strength, a Modulator Bias Controller (MBC) acting as a feedback loop to lock the modulator around its functioning point, and a photodiode used to monitor the output power and measure the average number of photons per coherent state $\langle n \rangle$ (which for the CV-QKD protocol is related to the modulation strength by $V_A = 2\langle n \rangle$).  A Digital-to-Analog Converter (DAC) connects the controlling computer to the different devices. The typical scheme for the source is presented in Figure \ref{fig:CVsource}. The energy cost is given by:

\begin{equation}
    S_{\mathrm{CV}} (t) = t(P_{\rm laser} + P_{\rm mbc} + P_{\rm dac} + P_{\rm photodiode}).
\end{equation}

\subsection{Manipulation}
\label{sec:fusion}
In this model, the manipulation of quantum states refers to all the electro-optical operations done to a photonic quantum state before detection. These manipulations are essential for almost all schemes since photonic states need to be shaped to perform the protocol or to control the measurement basis. When the information is encoded in the polarization of a photon, manipulation can be done using a motorized polarization controller before a PBS, with an energy cost given by:
\begin{align}
M_{\rm  polar}(t)= tP_{\rm modulator}.
\end{align}

Quantum gates for time bins are challenging since they require interferometry again. The energy cost is given by:
\begin{align}
M_{\rm time}(t)= t P_{\rm interferometry}.
\end{align}
\begin{figure}
    \begin{subfigure}{0.5\textwidth}
        \centering
        \includegraphics[width=0.5\textwidth]{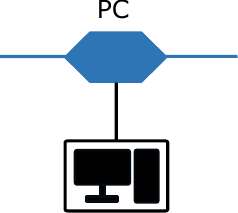}
        \caption{\raggedright\justifying Active manipulation of polarization-encoded states}
        \label{fig:DVpolmeasure}
    \end{subfigure}
    \begin{subfigure}{0.5\textwidth}
        \centering
        \includegraphics[width=0.5\textwidth]{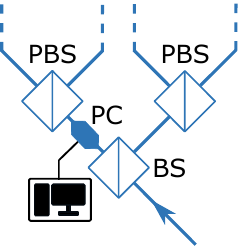}
        \caption{\raggedright\justifying Passive manipulation of polarization-encoded states}
        \label{fig:DVpolmeasure}
    \end{subfigure}
    \begin{subfigure}{0.5\textwidth}
        \centering
        \includegraphics[width=0.5\textwidth]{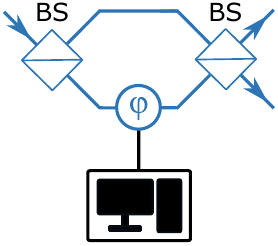}
        \caption{\raggedright\justifying Manipulation of time-encoded state}
        \label{fig:DVtimemeasure}
    \end{subfigure}
\caption{\raggedright\justifying \textbf{a)} Manipulation station for polarization qubits, which allows for projections over any polarization state.  \textbf{b)} Manipulation station for time-bin qubits.  It consists of an unbalanced Mach-Zehnder interferometer with phase control in one arm.}
\end{figure}

Some network protocols require multipartite entangled states, \textit{i.e.}, states with more than two entangled photons. To create such states, the most common technique is to join two bipartite sources together through an operation called \textit{fusion}~\cite{Browne_2005, Cao_2024,Fusion1, Pan2018}. This process takes one photon from each bipartite source and entangles them to create a four-photon entangled state. Higher orders are then reached by adding fusion stations and sources sequentially. Efficient fusion is challenging, especially with photonic states, since it requires precise timing of events that are inherently probabilistic. Recent progress towards fusion-based quantum computation has, however, shown that fusion of photonic graph-state can be the basis of quantum computers~\cite{bartolucci2021fusionbased}.\\

\begin{figure}
    \centering
    \includegraphics[width=0.45\textwidth]{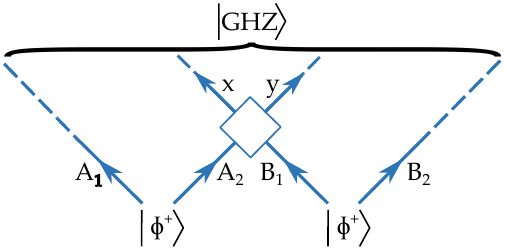}
    \caption{\raggedright\justifying Graphical representation of a generic fusion operation, one of the building blocks of multiparty entangled state creation.}
    \label{fig:FusionSchema}
\end{figure}

Figure \ref{fig:FusionSchema} details a graphical representation of the fusion operation that is considered in this work. Given input channels $A_2$ and $B_1$ and output channels $x$ and $y$, the logical operation required for fusion is defined as:
 \begin{equation}
 \begin{aligned}
 \ket{0}_{A_2} &\rightarrow \ket{0}_x,\\
 \ket{0}_{B_1} &\rightarrow \ket{0}_y,\\
 \ket{1}_{A_2} &\rightarrow \ket{1}_y,\\
 \ket{1}_{B_1} &\rightarrow \ket{1}_x,
 \end{aligned}
 \end{equation}
 
Indeed, acting on one qubit of two independent Bell pairs $\ket{\phi^+}_A = \frac{1}{\sqrt{2}}(\ket{0}_{A_1}\ket{0}_{A_2} + \ket{1}_{A_1}\ket{1}_{A_2})$ and $\ket{\phi^+}_B$ defined similarly, this logical operation gives:
\begin{align*}
\ket{\phi^+}_A \otimes \ket{\phi^+}_B 
&=\frac{1}{\sqrt{2}}\left(\ket{0000}_{A_1 A_2 B_1 B_2} + \ket{1100}_{A_1 A_2 B_1 B_2}\right. \\
&\hspace{0.6cm} +\left. \ket{0011}_{A_1 A_2 B_1 B_2} + \ket{1111}_{A_1 A_2 B_1 B_2}\right) \nonumber\\
&\rightarrow\frac{1}{\sqrt{2}} \left(\ket{0000}_{A_1 x y B_2} + \ket{1100}_{A_1 y y B_2}\right.\\ 
&\hspace{0.6cm} \left.+ \ket{0011}_{A_1 x x B_2} + \ket{1111}_{A_1 x y B_2}\right). \nonumber
\end{align*}
Assuming post-selection to keep only the states with one photon in each output channel, this eliminates the contributions from the two components in the middle, and leads to:
\begin{align}
    \frac{1}{\sqrt{2}}(\ket{0000}_{A_1 x y B_2}  + \ket{1111}_{A_1 x y B_2}),
\end{align}
which is a 4-photon GHZ state. Note that this post-selection process is necessary and that it causes the fusion process to be inherently probabilistic. The maximum success probability of photonic fusion achieved via this method is 50\%.  Higher fusion success probabilities can be reached by adding additional lasers and photonic ancillas~\cite{BSMwancilla1,BSMwancilla2}. \\

For polarization-encoded qubits, this operation can be done with a PBS. One of the inputs contains polarization compensating modulators, and therefore the energy cost is given by:
\begin{align}
M_{\rm fusion, polar}(t)= t P_{\rm modulator}. 
\end{align}

\begin{figure}
   \begin{subfigure}{0.45\textwidth}
        \centering
        \includegraphics[width=0.25\textwidth]{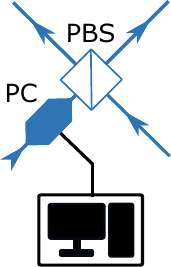}
        \caption{\raggedright\justifying Polarization fusion}
        \label{fig:DVpolfusion}
    \end{subfigure}
         \begin{subfigure}{0.45\textwidth}
        \centering
        \includegraphics[width=0.25\textwidth]{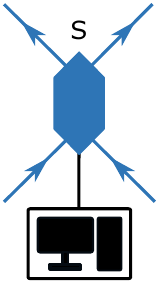}
        \caption{\raggedright\justifying Time fusion}
        \label{fig:DVtimefusion}
    \end{subfigure}

\caption{\raggedright\justifying (a) Fusion station for polarization qubits. The transformation required is that of a polarized beam splitter (PBS). (b) Fusion station for time bin entanglement, done by an intensity modulator.}
\label{fig:ECfusion}
\end{figure}

\begin{figure}
    \centering
    \begin{subfigure}{0.45\textwidth}
        \centering
        \includegraphics[width=0.65\textwidth]{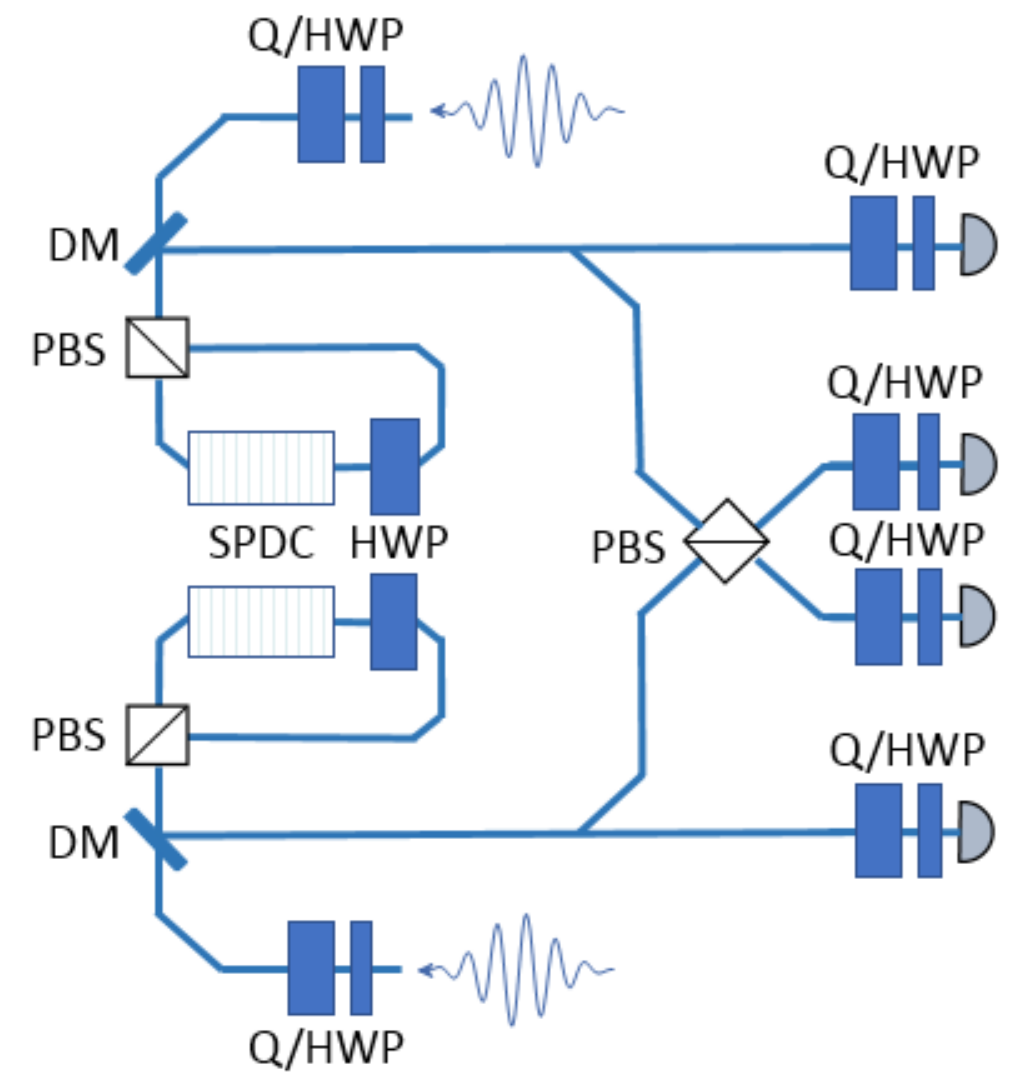}
        \caption{\raggedright\justifying GHZ in polarization}
    \end{subfigure}
         \begin{subfigure}{0.45\textwidth}
        \centering
        \includegraphics[width=0.65\textwidth]{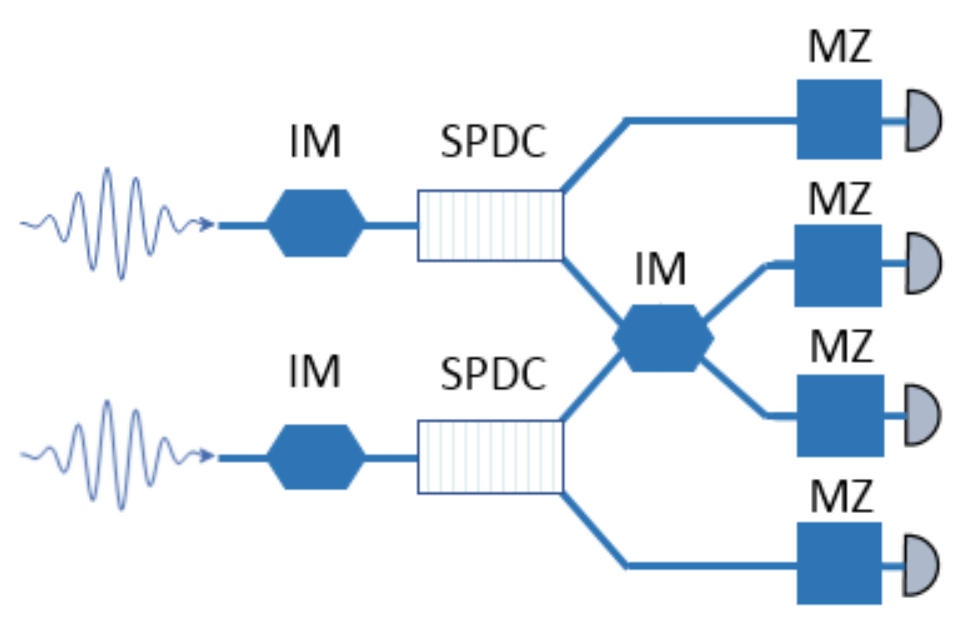}
        \caption{\raggedright\justifying GHZ in time}
    \end{subfigure}
    \caption{\raggedright\justifying Schematics of setups to create and share 4-qubit GHZ states encoded \textbf{(a)} in polarization and \textbf{(b)} in time.}
    \label{fig:ECGHZ}
\end{figure}

For time-encoded qubits, this operation requires an intensity modulator with two-inputs and two outputs that acts as a very fast switch. This device is connected to an accompanying waveform generator and the power consumption is:
\begin{align}
M_{\rm fusion, time}(t)= t P_{\rm modulator}.
\end{align}
The schematics of polarization and time fusion are shown in Figure \ref{fig:ECfusion}, while that of a complete setup to create 4-qubit GHZ states is shown in Figure \ref{fig:ECGHZ}.

\subsection{Detection}
\subsubsection{Single photon detectors}
In DV regimes, detection devices are typically threshold detectors, which are greatly influenced by the choice of signal wavelengths. Avalanche Photo-Diode (APD) single-photon detectors work through the ionization of their constituent material at the reception of photons, which creates a current that is in turn amplified. At near-infrared wavelengths, those are readily available and have good efficiencies and low energetic costs that make them an interesting choice (see Table \ref{tab:baselineparameters} and \ref{tab:tableofeverything}). Superconducting Nanowire Single-Photon Detectors (SNSPDs) have a much higher efficiency at telecom wavelength, as well as small jitters and dead times, allowing for high detection rates and precision. The interaction of the photons with the nanowire creates a temporary resistive region in the superconductive wire, briefly breaking superconductivity and leading to a detectable voltage pulse. SNSPDs are one of the leading choices in current experimental implementations of photonic protocols at telecom wavelength. However, the necessity of cryogenics presents a significant drawback in their energetic cost. \\

Incorporating essential electronics in the computer present at each node, such as Time-to-Digital Converters (TDC), two energy functions for the detectors can be formulated:
\begin{align}
D_{\rm APD}&=t  P_{\rm APD},\\
D_{\rm SNSPD}&= t  P_{\rm SNSPD}.
\end{align}

\subsubsection{Coherent detectors}

For CV protocols, detection can be either homodyne or heterodyne. In homodyne detection, a single quadrature of the field is measured. In heterodyne detection, both quadratures of the light field are measured simultaneously at the expense of the addition of extra noise in the signal.\\

The base device in both scenarios is the same: a Balanced Homodyne Detector (BHD) acts as the core component of the  apparatus~\cite{FundamentalsOfKikuch2016}.  It is internally composed of two standard photodiodes and a Trans-Impedance Amplifier (TIA) that transforms the current difference of the two photodiodes into a voltage with a significant gain. The energetic cost of this balanced receiver can be broken down as follows:

\begin{equation}
    P_{\rm BHD} = 2 P_{\rm photodiode} + P_{\rm TIA}
\end{equation}

\begin{figure}
    \begin{subfigure}[t]{0.5\textwidth}
        \centering
        \includegraphics[width=0.75\textwidth]{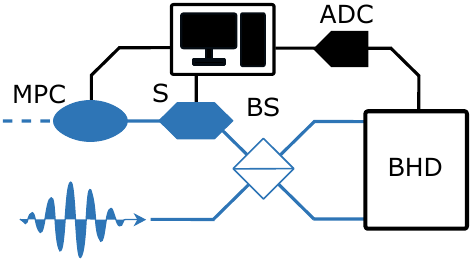}
        \caption{\raggedright\justifying Homodyne detection}
        \label{fig:CVhomodyne}
    \end{subfigure}
 \begin{subfigure}{0.5\textwidth}
        \centering
        \includegraphics[width=0.9\textwidth]{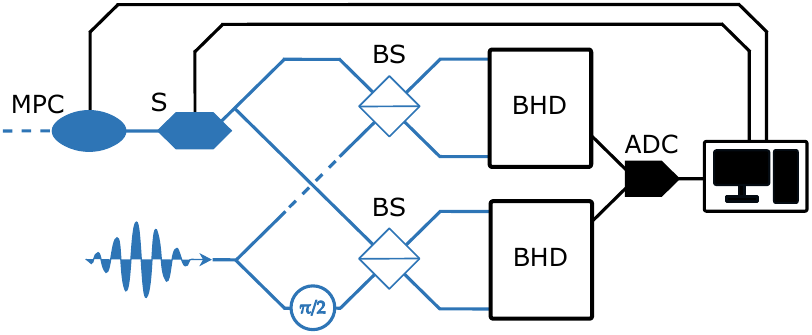}
        \caption{\raggedright\justifying Heterodyne detection}
        \label{fig:CVheterodyne}
    \end{subfigure}
    \caption{\raggedright\justifying Possible detection schemes for CV-QKD. MPC: Motorized Polarization Controller. S: Optical switch. BS: Beam Splitter. BHD: Balanced Homodyne Detector. ADC: Analog-to-Digital Converter.}
\end{figure}

For CV-QKD with homodyne detection, a Motorized Polarization Controller (MPC) and a Switch (S), used respectively to compensate long-distance polarization dispersion and to calibrate the noise levels, receive the light signal and mix it with a Local Oscillator (LO) in a beam splitter. The result is sent to the BHD device. It also requires a phase modulator on the LO path in order to select the measured quadrature (and perform a sifted protocol, as in BB84 for instance). An Analog-to-Digital Converter (ADC) allows the acquisition of data. For heterodyne detection, the general concept is the same, except that the signal light is divided in two. Each of these outputs is separately mixed with the local oscillator, one of the two being dephased by $\pi/2$, before being sent to a BHD each, allowing for simultaneous measurement of both quadratures. The two possible detection schemes are presented in Figures \ref{fig:CVhomodyne} and \ref{fig:CVheterodyne}.\\

The polarization controller can be avoided by performing a polarization-diverse protocol and performing the polarization compensation digitally. In that case, information can also be encoded on the second polarization. This requires adding a passive polarization beam splitter and a second detection station (with 1 or 2 BHDs depending on homodyne or heterodyne). The cost of the CV detection is given by:

\begin{equation}
    \begin{split}
        D_{\rm CV, hom, 1P}(t) &= t (P_{\rm  ADC} + P_{\rm laser} + P_{\rm BHD} + P_{\rm PC} + P_{\rm PM}),\\
        D_{\rm CV, het, 1P}(t) &= t (P_{\rm ADC} + P_{\rm laser} + 2 P_{\rm BHD} + P_{\rm PC}),\\
        D_{\rm CV, hom, 2P}(t) &= t (P_{\rm ADC} + P_{\rm laser} + 2 P_{\rm BHD}+P_{\rm PC} + P_{\rm PM}),\\
        D_{\rm CV, het, 2P}(t) &= t (P_{\rm ADC} + P_{\rm laser} + 4 P_{\rm BHD}+P_{\rm PC}),\\
    \end{split}
\end{equation}
where $1P$ and $2P$ correspond to single-polarization and double-polarization.

\subsection{Transfer of photons in fibers}
\label{sec:fiberloss}
A key component in quantum communications is the quantum channel linking the parties of a network. In this regard, optical fibers are the most common and stable components used to transfer photonic states. Fibers, however, come with the limitation that the probability that a photon is transmitted in the fiber decreases exponentially with the distance traveled. More specifically, the probability that a photon is transmitted after a distance $d$ (in km) in the fiber is given by the relation:
\begin{equation}
    \eta(d) = 10^{-d  \eta_{\rm fiber}/10},
\end{equation}
where $\eta_{\rm fiber}$ is the fiber loss coefficient, in dB/km, which depends on the wavelength of the photon going through. The different fibers loss coefficients that are considered for each wavelength are contained in Table \ref{tab:baselineparameters}.

\section{QKD protocols}
\label{app:QKDprotocols}
This section contains the model used to simulate each of the protocols' performances. The general setting in Quantum Key Distribution (QKD) is the following: two parties, Alice and Bob, wish to generate a common secret key. They are linked with a quantum and a classical channel. The quantum channel is used to transmit the quantum signals and is public, both for read and write access. The classical channel is also public but only for read access or, in other words, authenticated. \\

The scope of this study is focused on a few popular approaches, namely BB84, Entanglement-Based QKD (or E91), MDI-QKD, and Gaussian- and PSK-based CV-QKD. These protocols differ in performances and hardware involved, the details of which form the rest of this section. In theory, they all provide information-theoretic security. However, in practice, the implementation in a real quantum network can be subject to attacks. As they all provide different security properties, choosing one over another becomes a matter of context and implementation beyond simply energy consumption.

\subsection{Discrete Variable QKD}
\subsubsection{BB84}
\label{sec:BB84}
The protocol known as BB84~\cite{BB84} is the most straightforward approach to implement QKD. It was the first example of a protocol using the quantum properties of light to generate a secret key between two parties, ensuring information-theoretic security. In its simplest form, it requires a source of single photons, a way to encode qubits in different mutually unbiased bases, and a method to project the qubits on those bases that includes detectors. This gives considerable freedom in terms of component choices and implementations. Here, weak coherent states implementations~\cite{MadQCI, BB84coherent1} are studied as they are one of the most common implementations.\\

For the first part of the protocol, BB84 simply consists of sending photons from one party to another. The raw key rate of the protocol is thus given by the rate at which signals are sent by Alice and measured by Bob. As a simplification, we estimate the raw rate of BB84 as:

\begin{equation}\label{eq:Rate1photon}
    R_{\textrm{BB84}}=  \mu \; p_{\mathrm{coupling}} \; 10^{-\eta_{\mathrm{fiber}}\; L/10} \; p_{\mathrm{det}},
\end{equation}
where $r_{\mathrm{source}}$ is the repetition rate of the laser, $\mu$ is the mean photon number per pulse, $p_{\mathrm{coupling}}$ is the coupling probability into a fiber, $\eta_{\mathrm{fiber}}$ is the loss coefficient of the fiber used, $L$ the distance between the parties and $p_{\mathrm{det}}$ is the detection efficiency. These parameters depend on the set of hardware used and are the main variables in the simulations performed in this study.\\

The BB84 protocol involves one source, two motorized polarizing beam splitters to manipulate the states, and one detector station.  The classical cost $C_{\textrm{BB84}}(t)$ includes one computer for each party as well as a one-time tagger. These components are used to generate random bits to choose the creation and measurement bases of the photons, store the outcomes and perform sifting. The energy cost of a BB84 protocol is thus given by:

\begin{equation}\label{eq:BB84}
\begin{split}
E_{\textrm{BB84, polar/time}}(t) & = S_{\textrm{weak, polar/time}}(t)\\ &\hspace{0.4cm} + 2\;M_{\rm polar/time}(t)\\ &\hspace{0.5cm} +D_{\textrm{APD/SNSPD}}(t)\\ &\hspace{0.6cm} + C_{\textrm{BB84}}(t).
\end{split}
\end{equation}

\subsubsection{E91}
\label{sec:E91}
The protocol known as E91~\cite{Ekert} or entanglement-based QKD requires a source of entangled pairs of photons that are shared between Alice and Bob. By measuring photons arriving in random bases, Alice and Bob can extract a secret key. The raw rate at which the pairs are shared is: 
\begin{equation}
    \label{eq:E91rate}
    R_{\textrm{E91}}=\mu \; p_{\mathrm{coupling}}^2 \; 10^{\eta_{\mathrm{fiber}}\; L/10} \; p_{\mathrm{det}}^2,
\end{equation}
where $\mu$ is the probability of a successful Bell Pair generation by the SPDC process. Note in particular the factor $p_{\mathrm{det}}^2$ as a consequence of the fact that both parties need to measure a state. \\

In terms of hardware, a source of entangled single photon is required, and includes a laser and an oven. Both Alice and Bob must manipulate and detect the state. Two computers and two time-taggers are included in the classical components. The energy cost is given by:
\begin{equation}\label{eq:E91}
\begin{split}
E_{\textrm{E91, time/polar}}(t) & = S_{\textrm{SPDC, time/polar}}(t)\\&\hspace{0.4cm}+ 2 \; M_{\rm time/polar}(t) \\&\hspace{0.5cm}+ 2 \; D_{\textrm{APD/SNSPD}}(t)\\&\hspace{0.6cm}+ C_{\textrm{E91}}(t).
\end{split}
\end{equation}

\subsubsection{MDI-QKD}
Measurement Device Independent QKD~\cite{MDIQKD} is a scheme where the two parties, Alice and Bob, both produce a single photon and send it to a third party, Charlie, who performs a Bell-State Measurement (BSM). Despite the presence of a third party, the protocol is secure against eavesdroppers (or against a malicious Charlie) since correlations are measured instead of actual bit values.\\

We simplify the calculation of the raw rate of the MDI QKD protocol as follows. It is given by the rate at which two photons succeed to arrive simultaneously at the middle-station multiplied by the success probability of the BSM that is denoted as $p_{BSM}$. The following expression is obtained for the rate:

\begin{equation}
\label{eq:MDIQKDrate}
    R_{\textrm{MDI-QKD}}= \mu^2 \; p_{\mathrm{coupling}}^2 \; 10^{\eta_{\mathrm{fiber}}\; L/10} \; p_{BSM} \; p_{\rm det}^2 .
\end{equation}

A typical MDI-QKD setup therefore contains two sources of single photons, two ways to encode qubits, and a setup for the BSM in the chosen encoding that can have four detectors in one detection station. For polarization encoding, this operation can be done passively and therefore does not require energy except for the four required detectors, all present at one node. For time encoding, two interferometers are also necessary.
\begin{align}
M_{\rm BSM, polarization}(t)&= 0 \nonumber \\
M_{\rm BSM, time}(t)&= 2t P_{\rm interferometry}
\end{align}

Finally, a time-tagger and a computer are included in the classical components $C_{\textrm{MDI}}(t)$ for each party. The energy cost function becomes:

\begin{equation}
\begin{split}
E_{\textrm{MDI}}(t)&= 2\; S_{\textrm{weak,polar/time}}(t)\\&\hspace{0.4cm}+ M_{\rm BSM, polar/time}(t)\\&\hspace{0.5cm}+ D_{\textrm{APD/SNSPD}}(t)\\&\hspace{0.6cm}+ C_{\textrm{MDI}}(t).
\end{split}
\label{eq:MDIQKD}
\end{equation}

\subsubsection{Computation of the secret key rate of DV-QKD protocols}
The usual figure of merit for any QKD protocol is the \textit{secret key rate}, which depends not only on the rate at which photonic states are exchanged, but also on the noise affecting these states. For discrete variables, assuming that the QBER is the same in both measurement basis of the protocol, the upper bound on the secret key rate $K_{\rm DV-QKD}$ of BB84 and Entanglement-based protocols is given by the formula~\cite{shor2000simple,SecurityQKD,Watanabe_2007}:

\begin{equation}\label{eq:SKR}
    K_{\rm DV-QKD}= R\; (1 -2 \; h(QBER))
\end{equation}
where $R$ is the raw rate and $h$ is the binary entropy function given by $h(p)=-p\; \log_2(p) - (1-p)\log_2(1-p)$.\\

Note that this formula gives the maximal extractable secret key rate, in bit per channel use, from a given set of hardware components and a given noise. It assumes ideal post-processing, \textit{i.e.} error-correction and privacy amplification, while ignoring finite size effects. \\

In the case of MDI-QKD protocols, the computation of the secret key rate is more complex in general (see for example \cite{MDIQKD}). To simplify the comparison between approaches, we also use Equation \ref{eq:SKR} to extract the secret key rate of MDI-QKD.

\subsection{Continuous-Variable QKD}
\label{app:CVQKD}
\subsubsection{Definitions}
Continuous-Variable  QKD \cite{grosshans2002continuous,diamanti2015distributing,zhang2024continuousvariable} is based on employing infinite-dimensional quantum signals, typically coherent or squeezed states, to distribute secret keys. Modulated coherent states have the advantage of only requiring commercially available technologies, such as telecom lasers, balanced detectors and IQ modulators. 
In phase space, the set of possible points and their associated probabilities is called a constellation and represents the type of modulation that is being considered. For instance, for a Gaussian modulation, the two quadratures' average values follow Gaussian distributions. There also exist discrete modulations where the set of possible points is finite, and in CV-QKD, this can help for the error correction procedure. Possible discrete modulations are, for instance, $M$ Phase Shift Keying ($M$-PSK) where the $M$ points are uniformly distributed on a circle, $M$ Quadradure Amplitude Modulation ($M$-QAM) where the $M$ points are uniformly distributed on a grid, or QAM with Probabilistic Constellation Shaping such that $M$ points on the grid are associated to discretized Gaussian distributions~\cite{roumestan2022experimental}.\\

In this work, we study the energetic cost of CV-QKD with Gaussian modulated states, provided by variations of the GG02 protocol \cite{grosshans2002continuous,GaussianKeyRates_2009}, as well as the energetic cost of $4$-PSK, also called Quadratic Phase Shift Keying (QPSK) \cite{denys2021explicit}. In both cases, Alice generates and modulates coherent states of light to encode information and sends those state to Bob, who measures them using homodyne or heterodyne measurement. Alice and Bob end up with correlated variables that can be used to estimate channel parameters, and bound the information of an eavesdropper to derive a shared secret key. The energetic cost is given by:
\begin{equation}
\begin{split}
    E_{\rm CV, hom/het, 1P/2P}(t) &= S_{\rm CV}(t)\\&\hspace{0.4cm}+  D_{\rm CV, hom/het, 1P/2P}(t)\\&\hspace{0.6cm}+ C_{\rm CV}(t).
    \end{split}
\end{equation}

\subsubsection{Computation of the secret key rate of CV-QKD protocols\label{app:skr_cvqkd}}

In the asymptotic scenario, the secret key rate (in bits per symbol) of CV-QKD protocols $K_{\text{CV-QKD}}$ can be calculated with the Devetak-Winter formula \cite{devetak2005distillation}:
\begin{equation}\label{eq:CVRate}
    K_{\text{CV-QKD}} = \beta I_{AB} - \chi_{BE},
\end{equation}
where $\beta$ is the reconciliation efficiency, $I_{AB}$, the mutual information between Alice and Bob and $\chi_{BE}$, the Holevo bound on the information between Bob and Eve. As a particular distinction from discrete variable protocols, this scheme is crucially based on reverse reconciliation, where Alice adjusts her data to match Bob's raw key. \\

For this calculation, $V_A$ is the modulation strength chosen by Alice, which is twice the average photon number per symbol $V_A = 2\; \langle n \rangle$. $T$ is the transmittance of the channel, which can be related to the fiber distance via $T = 10^{-\frac{\eta d}{10}}$ where $d$ is the distance in km and $\eta$ the loss coefficient in dB/km.  $\xi$ is the excess noise of the channel (given at the input). $p_{\rm det}$ is the efficiency of the detection and $V_{el}$ is the electronic noise of every balanced detector. The value of $\beta = 95\%$ is assumed for the reconciliation efficiency, achievable with current Low Density Parity Check (LDPC) codes.\\

The first term in Equation~\eqref{eq:CVRate} can be computed from the capacity of the additive white Gaussian noise channel, given as:
\begin{equation}\label{eq:MInfoHom}
\begin{split}
    I_{AB, hom} ={}& \frac{1}{2}\log_2(1 +\text{SNR})\\ ={}& \frac{1}{2}\log_2\left(1 + \frac{p_{\rm det} T V_A}{1 + V_{el} + p_{\rm det} T \xi}\right),
    \end{split}
\end{equation}
for the homodyne scenario and as:
\begin{equation}\label{eq:MInfoHet}
\begin{split}
    I_{AB, het} ={}& \log_2(1 +\text{SNR})\\ ={}& \log_2\left(1 + \frac{p_{\rm det} T V_A}{2 + 2V_{el} + p_{\rm det} T \xi}\right),
\end{split}
\end{equation}
for the heterodyne case. The computation of Holevo's bound $\chi_{BE}$ is more involved. In the case of the Gaussian-modulated protocol, the model of~\cite{GaussianKeyRates_2009} is employed, which gives the bound:
\begin{equation}
    \chi_{BE} = \sum_{i=1}^2 G\left(\frac{\lambda_i-1}{2}\right) - \sum_{i=3}^5 G\left(\frac{\lambda_i-1}{2}\right).
\end{equation}
Here, $\lambda_1, \lambda_2$ are the symplectic eigenvalues of the covariance matrix characterizing the state shared between Alice and Bob before Bob's measurement and $\lambda_3, \lambda_4, \lambda_5$ are the symplectic eigenvalues of the covariance matrix characterizing the state shared by Alice and Bob after the homodyne or heterodyne detection. $G$ is the real function $G(x) = (x+1)\log_2(x+1) - x\log_2(x)$. By definition, $\lambda_5 = 1$ whereas the other symplectic eigenvalues are calculated according to the parameters of the implementation, provided the auxiliary parameters:
\begin{equation}
    \begin{split}
        \chi_{line} &= \frac{1}{T}-1+\xi,\\
        \chi_{hom} &= \frac{1-p_{\rm det}+V_{el}}{p_{\rm det}},\\
        \chi_{het} &= \frac{1+(1-p_{\rm det})+2V_{el}}{p_{\rm det}},\\
        \chi_{tot, hom/het} &= \chi_{line} + \frac{\chi_{hom/het}}{T},\\
        V &= V_A+1.
    \end{split}
\end{equation}
One can compute $\{\lambda_i\}_{i=1}^4$ as:
\begin{equation}
\begin{split}
    \lambda^2_{1,2} &= \frac{1}{2}\left[A\pm\sqrt{A^2-4B}\right],\\
    A &= V^2(1-2T)+2T+T^2(V+\chi_{line})^2,\\
    B &= T^2(V\chi_{line}+1)^2,
\end{split}
\end{equation}
and: 
\begin{equation}
\begin{split}
    \lambda^2_{3,4} &= \frac{1}{2}\left[C\pm\sqrt{C^2-4D}\right],\\
    C_{hom} &= \frac{A\chi_{hom} + V\sqrt{B}+T(V+\chi_{line})}{T(V+\chi_{tot,hom})},\\
    D_{hom} &= \sqrt{B}\frac{V+\sqrt{B}\chi_{hom}}{T(V+\chi_{tot,hom})},\\
    C_{het} &= \frac{1}{(V(T+\chi_{tot, het}))^2}\left[A\chi_{het}^2+B+1\right.\\
    &\left.+2\chi_{het}(V\sqrt{B}+T(V+\chi_{line}))+2T(V^2-1)\right],\\
    D_{het} &= \left(\frac{V+\sqrt{B}\chi_{het}}{T(V+\chi_{tot,het})}\right)^2.
\end{split}
\end{equation}
For the PSK modulation, the analysis of \cite{denys2021explicit} is used as a guide. Using again Equation~\eqref{eq:MInfoHom} or Equation~\eqref{eq:MInfoHet} for the mutual information according to the measurements, the Holevo bound can be computed with the following quantities:
\begin{equation}
    \begin{split}
        V &= V_A+1,\\
        W &= 1+p_{\rm det} T V_A + p_{\rm det} T \xi +V_{el},\\
        Z &= \sqrt{T}\left(2\alpha^2 e^{-\alpha^2}\sum_{k=0}^{M-1}\frac{\nu_k^{3/2}}{\nu_{k+1}^{1/2}}\right.\\
        &- \left.\sqrt{2\xi\alpha^2}\sqrt{e^{-\alpha^2}\sum_{j=0}^{M-1}\frac{\nu_j^2}{\nu_{j+1}} - e^{-2\alpha^2}\left(\sum_{j=0}^{M-1}\frac{\nu_j^{3/2}}{\nu_{j+1}^{1/2}}\right)^2} \right),
    \end{split}
\end{equation}
where: 
\begin{equation}
    \begin{split}
    \nu_k &= \frac{1}{M}\sum_{j=0}^{M-1}e^{-ijk\frac{2\pi}{M}}\exp\left(\alpha^2e^{ij\frac{2\pi}{M}}\right),\\
    \alpha^2 &= \frac{V_A}{2}.\\
    \end{split}
\end{equation}
One can then consider the covariance matrix 
\begin{equation}
    \Gamma = \begin{pmatrix}
        V\mathbb{I}_2 & Z\sigma_Z\\
        Z\sigma_Z &W\mathbb{I}_2\\
    \end{pmatrix}
\end{equation}
where $\mathbb{I}_2$ is the 2x2 identity matrix and $\sigma_Z$ is the Pauli Z matrix. The value of the Holevo bound is then given by:
\begin{equation}
    \chi_{BE} = G\left(\frac{\lambda_1-1}{2}\right)+G\left(\frac{\lambda_2-1}{2}\right)-G\left(\frac{\lambda_3-1}{2}\right),
\end{equation}
where $\lambda_1$ and $\lambda_2$ are the symplectic eigenvalues of $\Gamma$. $\lambda_3$ is either $\lambda_3 = V - \frac{Z^2}{W+1}$ for homodyne detection or $\lambda_3 = \sqrt{V(V-\frac{Z^2}{W})}$ for heterodyne detection. Note that the formula for $Z$ was given for a general $M$-PSK scenario, and setting $M=4$ gives the results for a QPSK modulation.

\subsubsection{Computation of the secret key rate of the CV-CKA protocol}
\label{app:SKRcvcka}
Regarding the CV-CKA protocol, a brief description of the full model provided in \cite{ottaviani_modular_2019} follows, composed of multiple Bobs. There, each Bob (separated equidistantly) employ coherent states with a Gaussian modulation, such that the secret key rate is fully determined by covariance matrices. For every round, the generated states are sent to an untrusted relay, which performs diverse Bell measurements through a series of beam splitters and homodyne detectors. Following the notation of said reference, the modulation of the initial states as is denoted as $\mu$, the thermal noise as $\delta = (1-p_{\rm det}+V_{el})/p_{\rm det}$ and $\omega = 2\delta+1$. The relevant quantity here is then the covariance matrix shared by any two Bobs $i$ and $j$ after the Bell measurements
\begin{equation}
    V'_{B_i B_j} = \begin{pmatrix}
        \Delta & \Theta \\
        \Theta & \Delta
    \end{pmatrix}.
\end{equation}
Here: 
\begin{align*}
    \Delta &= \mathrm{diag}\{y - (n-1)z^2/(n x),y - z^2/(n x) \}, \\ \Theta &= \mathrm{diag}\{z^2/(n x),- z^2/(n x) \},
\end{align*}
where $N$ is the number of users, and:
\begin{align}
    x &= T \mu + (1-T \mu), \omega  \nonumber \\
    y &= \mu,\\
    z &= \sqrt{T (\mu^2 -1)}. \nonumber 
\end{align}
The mutual information between the Bobs is then:
\begin{equation}\label{eq:MInfoCVCKA}
    I_{B_i B_j} = \frac{1}{2}\log \left(\frac{1 + \mathrm{det}(V'_{B_i}) + \mathrm{tr}(V'_{B_i})}{1 + \mathrm{det}(V''_{B_j}) + \mathrm{tr}(V''_{B_j}) } \right),
\end{equation}
where $V''_{B_j}$ denotes the covariance matrix of one the Bobs after the other has performed a homodyne measurement. On similar grounds, the relevant Holevo information is:
\begin{equation}\label{eq:ChiCVCKA}
    \chi_{B_i E} = 2 G\left( \frac{\nu-1}{2}\right) - G\left( \frac{\nu_n-1}{2}\right),
\end{equation}
where:
\begin{align*}
    \nu = \sqrt{y \left(y - \frac{z^2}{x}\right)}, && \nu_n = \sqrt{\frac{\lambda \bar\lambda}{\tau \bar{\tau}}},
\end{align*}
with:
\begin{align}
\begin{split}
    \lambda &= n \omega \mu + T [1 + (n-1-n\omega)\mu],   \\
    \bar{\lambda} &= n \omega \mu + T [n - 1 - (n\omega -1)\mu], \\
    \tau &= n \omega (1-T) + T (n-1+\mu),   \\
    \bar{\tau} &= n \omega (1-T) + T [(n-1)\mu + 1]. 
\end{split}
\end{align}
Inserting both Equation~\eqref{eq:MInfoCVCKA} and Equation~\eqref{eq:ChiCVCKA} in Equation~\eqref{eq:CVRate}, the secret key rate is obtained.

\section{Finite-Size effects in CV QKD}\label{app:fse}

\begin{figure}[h]
    \centering
    \includegraphics[width=\linewidth]{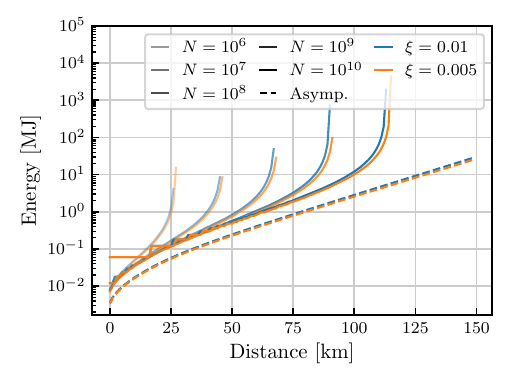}
    \caption{Finite size energetic cost of CV-QKD protocol. The protocol here uses single-polarization heterodyne detection. The DSP cost is set to $\tau_{\rm DSP} = 6$ mJ/Symbol.}
    \label{fig:cvqkd-fse}
\end{figure}

In general, the asymptotic key rate is not sufficient as a finite number of states are exchanged between Alice and Bob.
For CV-QKD, we consider blocks of $N$ symbols. The goal is still to find the time $t$ that is required for the protocol to run given a target number of bits $N_{\rm target}$. We consider that the finite-size key rate, for the Gaussian modulation, is given by 

\begin{equation}
    K_{\rm FSE} = \frac{n}{N}(1-{\rm FER})(\beta I_{AB} - \chi_{BE}^{\varepsilon_{\rm PE}} -\Delta (n))
\end{equation}

following the analysis of~\cite{PhysRevA.81.062343}, where $n$ is the number of symbols used for the key generate ($m=N-n$ are used for parameter estimation), $\rm FER$ is the frame error rate, $\chi_{BE}^{\rm \varepsilon_{\rm PE}}$ is the Holevo bound compatible with the observation of the parameters, up to an error probability $\varepsilon_{\rm PE}$ and $\Delta(n)$ a correction term related to the security of privacy amplification. The value of $\Delta(n)$ is 

\begin{equation}
    \Delta (n) = (2\dim(\mathcal{H}_X)+3)\sqrt{\frac{\log_2(2/\bar{\varepsilon})}{n}} +\frac{2}{n}\log_2(1/\varepsilon_{\rm PA})
\end{equation}

where $\mathcal{H}_X$ is the Hilbert space used in the raw key, $\bar{\varepsilon}$ a smoothing parameter and $\varepsilon_{\rm PA}$ is a free parameter that arises from the left-over has lemma.

As for, $\chi_{BE}^{\rm \varepsilon_{\rm PE}}$, it can be computed by using worst case estimators:

\begin{equation}
    \begin{split}
        t_{\rm min} &= \sqrt{\eta T} - z_{\varepsilon_{\rm PE}/2}\sqrt{\frac{1+\eta T \xi}{m V_A}}\\
        \sigma^2_{\rm max}&= 2+2V_{\rm el}+\eta T\xi +z_{\varepsilon_{\rm PE}/2}\frac{\sigma^2\sqrt{2}}{\sqrt{m}}\\
    \end{split}
\end{equation}

with $t$ an estimator of $\sqrt{T}$ and $\sigma^2$ an estimator of the overall variance of Bob. The overall security parameter is given by

\begin{equation}
    \varepsilon = \varepsilon_{\rm PE} + \varepsilon_{\rm EC} + \bar{\varepsilon} +\varepsilon_{\rm PA}
\end{equation}

As in~\cite{PhysRevA.81.062343}, we take $\varepsilon_{\rm PE} = \varepsilon_{\rm EC} = \bar{\varepsilon} =\varepsilon_{\rm PA} = 10^{-10}$.

The goal is to use a certain number of frame $k$ that satisfies $k\cdot N\cdot K_{\rm FSE} \geq N_{\rm target}$, which is $k = \left\lceil \frac{N_{\rm target}}{N \cdot K_{\rm FSE}}\right\rceil$, hence the time to run the protocol is

\begin{equation}
    t = \frac{kN}{r_{\rm source}} = \frac{ \left\lceil \frac{N_{\rm target}}{N \cdot K_{\rm FSE}}\right\rceil N}{r_{\rm source}}
\end{equation}

Taking into account the cost of DSP and the initialization cost, one finds

\begin{widetext}
\begin{equation}
    E_{\rm CV-QKD}^{\rm FSE} = E_{\pi}(0) + \left\lceil \frac{N_{\rm target}}{n(1-{\rm FER})(\beta I_{AB} - \chi_{\rm BE}^{\varepsilon_{\rm PE}}-\Delta(n))}\right\rceil N\left(\tau_{\rm DSP} + \frac{P_{\pi}}{r_{\rm source}}\right)
\end{equation}
\end{widetext}


This quantity is plotted in Fig.~\ref{fig:cvqkd-fse}, with $m = \frac{N}{2}$ and otherwise the same parameters as before for different values of $N$. The asymptotic value is also reminded in dashed lines.

\section{Time-bin encoding}\label{app:timebin}
\begin{figure}[!ht]
    \centering
    \includegraphics{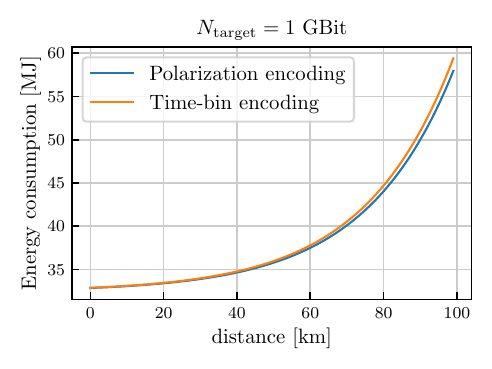}
    \caption{\raggedright\justifying  Energy required to distill 1Gbit of secret key using different choices of encoding.}
    \label{fig:BB84timebin}
\end{figure}

To grasp the influence of the choice of encoding, polarization and time encoding are compared. In Figure \ref{fig:BB84timebin} is shown the theoretical energy necessary to obtain 1 Gbit of secret key using BB84 between two parties as a function of the distance for a fixed QBER of 1\%. In terms of energy consumption, using a time-bin based setup amounts to the addition of a modulator to carve the pulses into bins in time, while a polarization based setup includes electro-optic modulators to select the polarization. The influence of this choice of encoding on the energy consumption is relatively small despite the requirements of interferometry for time encoded protocols. This small difference could, however, prove to become relevant when the network scales up. \\

The same result holds for the other DV protocols considered in this study: using time-bin encoding results in a slight increase in the energy consumption. Due to the lack of major differences, time encoded plots for QKD and CKA were excluded from Figures \ref{fig:EE}, \ref{fig:DVprotocols} and \ref{fig:CKAstudyfull} for readability.

It is important to note that the rates of all protocols are related to the chosen hardware: if available, it could be advantageous to select faster devices, but these can come at vastly different energy costs.
 
\section{Comparison with measured values}
\label{app:measured}

\begin{figure}[!h]
    \centering
    \includegraphics[scale=1]{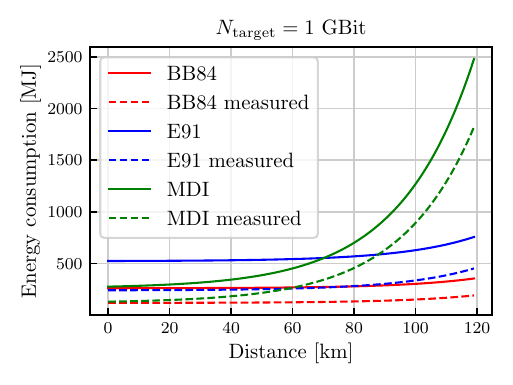}
    \caption{\raggedright\justifying Comparison between the theoretical (plain lines) and the measured (dashed lines) values of the energy required to distill 1 Gbit of secret key using the three DV-QKD protocols. Note that the detector that we could study to produce this plot have an energy consumption of 3000W.}
    \label{fig:DVmeasured}
\end{figure}

As explained with Table \ref{tab:tableofeverything}, some values for the energy consumption of the hardware elements used in this study have been measured directly in a lab. In Figure \ref{fig:DVmeasured}, the difference on the energy required to distill 1 Gbit of secret key using the three DV-QKD protocols studied in this work is shown when using both the measured values and the theoretical values. Since the measured values are almost always lower than the theoretical ones, the overall energy consumption is also lower, as expected. This is coherent with the fact that hardware manuals give an upper bound on the energy consumption. The real consumption of protocols is thus lower than the predictions, although the order of magnitude remains correct.

\newpage
\section{Distribution of power consumption}
\label{app:distribution}
\begin{figure}[b]
    \centering
    \includegraphics[scale=0.85]{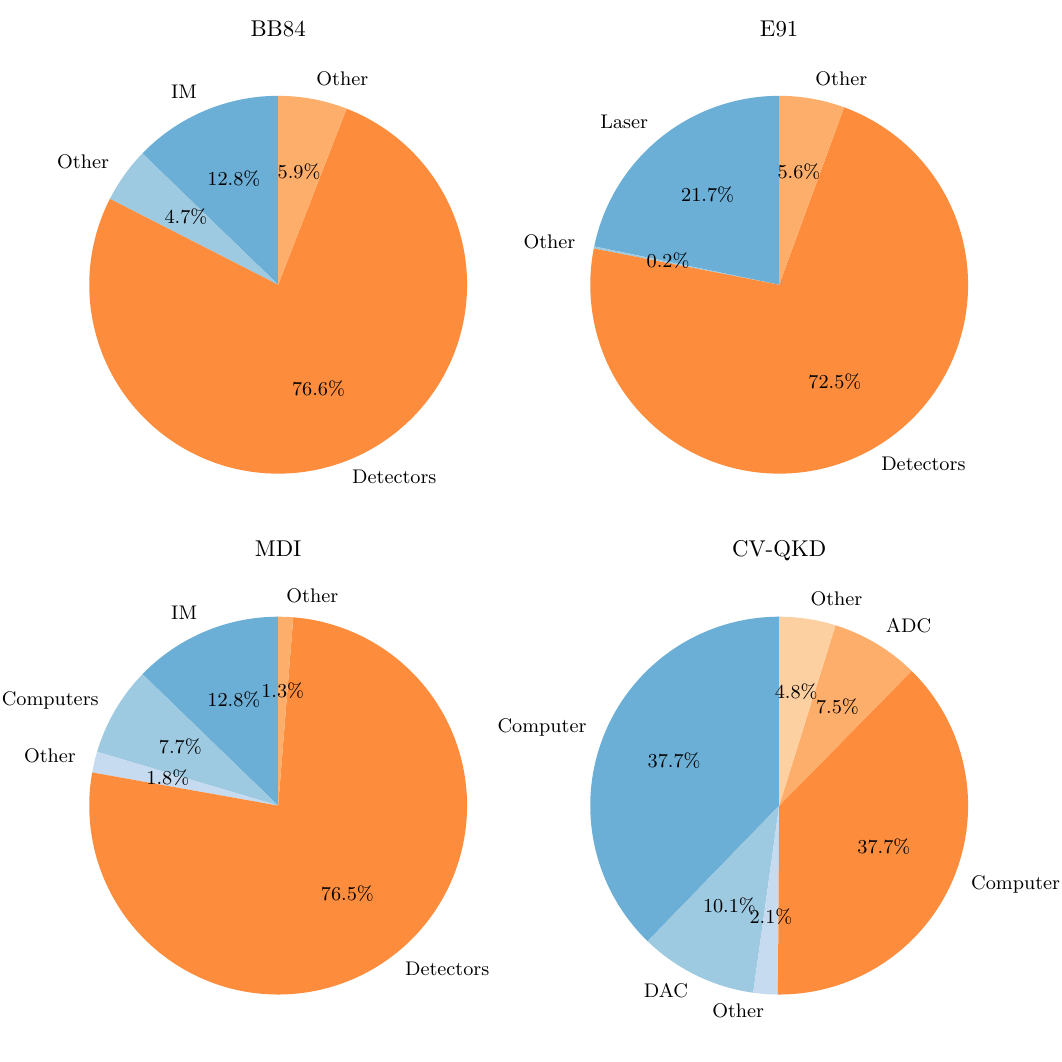}
    \caption{\raggedright\justifying Distribution of power consumption for the different QKD protocols. Shades of blue are used for the source components and shades of orange for the detection components. A darker color indicates a higher power contribution.}
    \label{fig:piecharts}
\end{figure}

In Figure \ref{fig:piecharts}, the distribution of power usage between components for the studied QKD protocols (BB84, E91, MDI and CV) is plotted as a pie charts, using shades of blue for the source and shades of orange for the detection.  \\

Clearly, the biggest contribution for the DV protocols is the detection with the SNSPDs occupying around $75\%$ of the power consumption. For E91, the laser is then the second biggest source of energy consumption, which is due to the high energy required for the generation of photon pairs through non-linear effects. For CV-QKD, the distribution between Alice and Bob is almost equal with the biggest contribution coming from the computers and then from the DAC and ADC.

\onecolumngrid

\twocolumngrid


\bibliographystyle{IEEEtran}
\bibliography{biblio}

\bibliographystyleH{IEEEtran}
\bibliographyH{biblio}

\end{document}